\documentclass[12pt,preprint]{aastex}
\shorttitle{SN~2006bp}
\shortauthors{Quimby et al.}
\usepackage{epsfig}
\begin{document}

\title{SN~2006bp: Probing the Shock Breakout of a Type II-P Supernova\footnotemark[1]}

\author{
  Robert M. Quimby\altaffilmark{2},
  J. Craig Wheeler\altaffilmark{2},
  Peter H\"{o}flich\altaffilmark{3},
  Carl W. Akerlof\altaffilmark{4},
  Peter J. Brown\altaffilmark{5}, and
  Eli S. Rykoff\altaffilmark{4} 
}
\footnotetext[1]{Based on observations obtained with the Hobby-Eberly
Telescope, which is a joint project of the University of Texas at
Austin, the Pennsylvania State University, Stanford University,
Ludwig-Maximilians-Universit\"{a}t M\"{u}nchen, and
Georg-August-Universit\"{a}t G\"{o}ttingen.}
\altaffiltext{2}{Department of Astronomy, University of Texas, Austin, TX 78712, USA}
\altaffiltext{3}{Department of Physics, Florida State University, Tallahassee, FL 32312, USA}
\altaffiltext{4}{University of Michigan, 2477 Randall Laboratory, 450
                 Church St., Ann Arbor, MI, 48104, USA}
\altaffiltext{5}{Astronomy \& Astrophysics Department, The Pennsylvania State University, University Park, PA 16802, USA}

\begin{abstract}

HET optical spectroscopy and unfiltered ROTSE-III photometry spanning
the first 11 months since explosion of the Type II-P SN~2006bp are
presented. Flux limits from the days before discovery combined with
the initial rapid brightening suggest the supernova was first detected
just hours after shock breakout. Optical spectra obtained about 2 days
after breakout exhibit narrow emission lines corresponding to
\ion{He}{2} $\lambda$4200, \ion{He}{2} $\lambda$4686, and \ion{C}{4}
$\lambda\lambda$5805 in the rest frame, and these features persist in
a second observation obtained 5 hours later; however, these emission
lines are not detected the following night nor in subsequent
observations. We suggest that these lines emanate from material close
to the explosion site, possibly in the outer layers of the progenitor
that have been ionized by the high energy photons released at shock
breakout.
A P-Cygni profile is observed around 4450~\AA\ in the $+2$ and $+3$
day spectra. Previous studies have attributed this feature to high
velocity H$\beta$, but we discuss the possibility that this profile is
instead due to \ion{He}{2} $\lambda$4687. Further HET observations (14
nights in total) covering the spectral evolution across the
photometric plateau up to 73 days after breakout and during the
nebular phase around day $+340$ are presented, and expansion velocities
are derived for key features. The ROTSE-III light curve shows that the
plateau phase lasted until about day $+80$ and the exponential decay
phase began around 110 days after breakout. The measured decay slope
for the unfiltered light curve is $0.0073 \pm 0.0004$~mag~day$^{-1}$
between days $+121$ and $+335$, which is significantly slower than the
decay of rate $^{56}$Co. We combine our HET measurements with published
X-ray, UV, and optical data to obtain a quasi-bolometric light curve
through day +60. We see a slow cooling over the first 25 days, but no
sign of an early sharp peak; any such feature from the shock breakout
must have lasted less than $\sim 1$ day.

\end{abstract}


\section{Introduction}\label{intro}

Type II supernovae (SNe~II) are explosions marking the deaths of
massive stars with significant amounts of hydrogen still intact. They
do not typically obtain the high peak luminosities nor the uniformity
exhibited therein by their thermonuclear cousins, the Type Ia events;
however, what SNe~II lack as intrinsic standard candles is compensated
for by their relative limpidity: SNe~II are easier to model as the
explosions do not involve steep chemical gradients, and the
progenitors of a few events have been identified (see \citealt{li2007}
for a recent example). SNe~II hold the promise of serving as
independent and perhaps absolute distance indicators
\citep{kirshner_kwan1974,shaviv85,hoeflich1991,schmidt94,hamuy_pinto2002,baron2004,nugent2006}.
In recent years, several Type II-P supernovae (SNe~II-P), where the
``P'' indicates a plateau phase in the light curve, that were
discovered at early times have been observed and modeled in detail
(SN~1999em;
\citealt{baron2000,hamuy2001,leonard2002,elmhamdi2003,baron2004,baklanov_blinnikov2005,dessart_hillier2006};
SN~1999gi; \citealt{schlegel2001,leonard2002b}; SN~2005cs;
\citealt{pastorello2006,tsvetkov2006,brown2006,baron2007}).
Polarization observations of SNeII-P have shown the outer,
Hydrogen-rich envelope to be rather spherical and that the
polarization increases at late times, which can be understood in terms
of asymmetric excitation \citep{leonard03,hoeflich01b}.

SNe~II-P are thought to result from stars massive enough to ignite
Carbon ($>$ 7-8 $M_\odot$; \citealt{eldridge_tout2004}), but light
enough to avoid envelope-stripping winds ($<25 M_\odot$;
\citealt{heger2003}). These explosions begin after the energy release
from nuclear activity in the core ceases to provide sufficient
pressure to support the stellar interior, and the core collapses under
its own weight, most likely forming a neutron star. The processes by
which the energy released in the collapse subsequently unbinds the
envelope are still a matter of debate \citep{khokhlov99}, but neutrino
interactions \citep{janka2006}, shock instabilities
\citep{blondin_mezzacappa2006}, acoustic instabilities
\citep{burrows2006}, and magneto-rotational effects
\citep{akiyama2003} may each play a role.

Gravitational waves are emitted during the bounce back phase and
neutrinos escape moments later, but the electromagnetic signal is
initially swallowed by the envelope. An outside observer would not
detect any change in the light output by the star until the
hydrodynamic front or the light front reached the photosphere, which
could be up to 3 days after the explosion depending on the extent of
the progenitor's envelope. The ensuing phase is known as the shock
breakout and it is manifest by a sudden, rapid increase in the
radiated luminosity. The emergent spectrum will have its peak energy
in the X-ray to UV bands, and this hard emission may ionize any
circumstellar material or recent wind. The bolometric light curve will
peak hours to days after breakout before decreasing in an adiabatic
free expansion phase to the plateau stage. During the plateau, a
cooling recombination wave recedes (in the mass frame) through the
ejecta layers, and the balance of a slowly increasing photospheric
radius with an effective temperature tied closely to the recombination
temperature will conspire to maintain the nearly constant plateau
luminosity. After the photosphere has swept through the bulk of the
envelope, a nebular phase ensues and the light curve fades
exponentially, now powered predominantly by the decay of $^{56}$Co.

The generic observational consequences of a shock breakout were laid
out by \citet{falk_arnett1977}. Thus far the only observations
acquired during the breakout phase were for SN~1987A (a peculiar Type
II), SN~1993J (a peculiar Type II/Ib hybrid), and SN~2006aj (a
peculiar Type Ic related to GRB 060218), and the bulk of light curve
modeling in the past two decades has been focused on SN~1987A or
models without extended photospheres or winds
\citep{litvinova85,arnett89,woosley88,hoeflich1991,ensman_burrows1992,eastman94,blinnikov00}.
While there is now good agreement found for the observations and
models of the shock breaking out of SN~1987A's compact progenitor,
little progress has been made in the expected properties of breakout
in an extended envelope as expected for common SNeII-P. In particular,
detailed, high resolution calculations of the shock breakout including
the hydrodynamical, ionization, and light fronts expected in red
supergiant atmospheres are still missing
\citep{mair92,chieffi03}. Polarization observations indicate that
differential runtime effects within the extended progenitor envelope
are likely to be small (i.e. the shock front reaches the photosphere
at a uniform time; \citealt{leonard03,hoeflich01b}). As illustrated by
\citet{falk_arnett1977}, the properties of the breakout peak depend on
the envelope mass and especially the density structure of the outer
layers. We may therefore gain a better understanding of the envelope
structure of massive stars in their final years by studying the early
time light curves and spectra of SNeII-P.

An astute amateur astronomer, Koichi Itagaki of Yamagata, Japan, first
identified SN~2006bp on unfiltered images of NGC 3953 taken with an
0.60-m f/5.7 reflector around April 9.60 UT
\citep{itagaki2006}. Mr. Itagaki observed the SN to brighten rapidly
(about 0.9 mag in 4 hours), and he quickly made the discovery public
via IAU Circular 8700 (submitted by S. Nakano). The host galaxy was
observed frequently in the days prior to discovery by our Texas
Supernova Search (Quimby et al. in prep.), resulting in pre-discovery
detections and stringent limits. Observations by the {\it Swift}
satellite beginning April 10.54 revealed a fading X-ray source
coincident with the explosion site, while observations by the VLA put
an upper limit to the 22.46 GHz flux of $< 0.4$ mJy on April 11.97
\citep{immler_brown2006,kelley2006,immler2007}. We obtained an optical
spectrum with the Hobby-Eberly Telescope (HET) on April 11.11, which
showed SN~2006bp to be an early Type II event~\citep{quimby2006b}.

In this paper we present photometric and spectroscopic observations
and analysis of SN~2006bp. We begin with a description of the
unfiltered optical ROTSE-III (Robotic Optical Transient Search
Experiment) observations in \S\ref{phot} including flux upper-limits
in the days leading up to shock breakout. The complete light curve is
presented and analyzed in \S\ref{lc} leading to an estimate of the
shock breakout date. The spectral observations from the HET are
detailed in section \ref{spec}. Data from the first two nights are
discussed in \S\ref{early_spec}, and the full spectral evolution is
considered in \S\ref{line_evol}. We give an approximate reconstruction
of the early-time quasi-bolometric light curve in
\S\ref{bolo}. Discussion and conclusions are presented in
\S\ref{conclusions}.

\section{Photometry}\label{phot}

We detected SN~2006bp as part of the Texas Supernova Search's nightly
monitoring of the Ursa Major Galaxy Cluster (Quimby et al. in prep.)
with the 0.45-m f/1.9 ROTSE-IIIb telescope \citep{akerlof03} at the
McDonald Observatory. The ROTSE-III telescopes gather unfiltered
optical light; the approximate spectral response is given in Table
\ref{rotse_response}. The SN is located in the overlap of three of our
search fields, and typically two of these fields were covered in the
nights leading up to the discovery. On April 9.15, the SN was formally
detected by the image subtraction pipeline at the $4.8\sigma$ level in
both of the fields covered that night, but these detections were still
below the $5\sigma$ threshold employed by the automatic candidate
filtering procedure used to limit spurious events, and so the SN was
not identified as a possible supernova until the following
night. Beginning on April 11, we obtained additional unfiltered images
centered on SN~2006bp with both ROTSE-IIIb and ROTSE-IIId, which is
located at the Turkish National Observatory at Bakirlitepe,
Turkey. All images were flat fielded and dark subtracted by the
standard ROTSE-III image processing pipeline \citep{rykoff_phd}. The
ROTSE-III instruments have a $1\fdg85 \times 1\fdg85$ field of view;
however, the PSF varies significantly across the frame, and the
instrumental zeropoint can as well under non-photometric
conditions. We therefore selected $30' \times 30'$ sub-frames centered
on the SN for the final photometry.

\begin{figure}
\epsscale{1.0}
\plotone{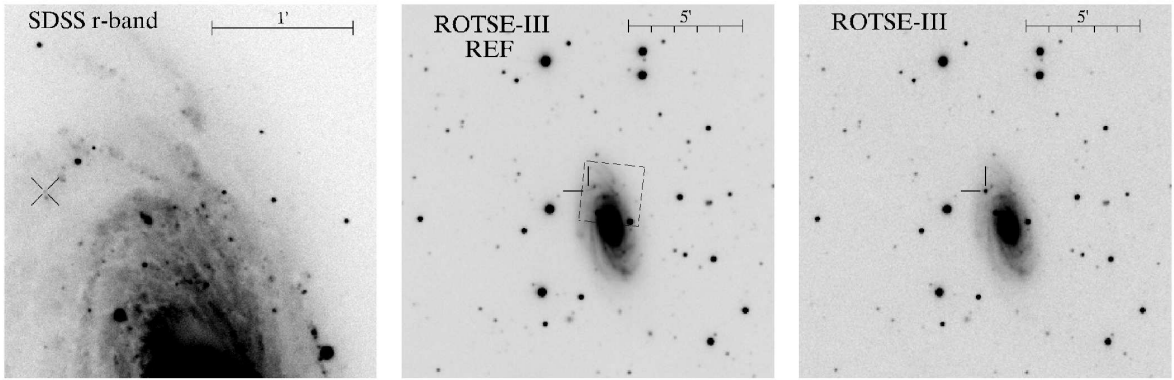}
\caption{Finding chart for SN~2006bp. The middle image shows the
ROTSE-III reference template, which was constructed from
post-explosion data as explained in the text. The box marks the
overlap with the detail of the SDSS r-band data shown in the left
image. The location of the SN is marked with an ``X''. The image on
the right shows the SN as it appeared in ROTSE-IIIb data from May 23
(near the middle of the plateau). North is up and east is to the left
in each image.}
\label{finder}
\end{figure}

Although we have a number of reference images taken prior to the
explosion of SN~2006bp in each of the three search fields, in each
case the location of the SN appears near the edge of the field where
the image quality is at its worst, and there is very little overlap
between the fields. Lacking a quality reference image, we decided to
construct a reference template using the best post-explosion images
available following the general procedure of \citet{li2003}. Briefly,
we co-added 44 ROTSE-IIIb images acquired at various phases of the
explosion, empirically measured the PSF, and then subtracted the
scaled PSF to remove the SN light. We were able to test the resulting
template by subtracting it from the images taken prior to the SN
explosion. No significant positive or negative residual was found on
these subtracted frames, which validates the template (hereafter
``REF''). Because the PSF on the REF is superior (much sharper) than
is obtained from the pre-explosion images, this greatly improves our
ability to convolve the template to match the delivered PSF on a given
night, and we are able to use the same template for all of the
available images. This allows for better and more consistent
measurements of the SN flux on the template subtracted frames. The
template is shown in Figure \ref{finder}.

To determine the flux evolution of SN~2006bp, we first made separate
co-additions for each unique night/field/telescope combination. In
doing so, we only used the best images available, defined to be those
images with detections for at least 80\% of all the objects found on
the deepest frame in that set. We median filtered the individual
images and masked bad pixels to create the coadded image (hereafter
``NEW''). If any of the pixels within 1-FWHM of the SN position were
masked, the image was removed from the NEW.

We performed PSF matched image subtraction of the REF from the NEWs
using a modified version of the Supernova Cosmology Project's code. A
selection of isolated, point-like objects were chosen to serve as
field standards (REFSTARS). The local PSF was first measured
empirically on the REF and on each NEW using the REFSTARS, and this
template was then fit to each REFSTAR to determine the PSF-fit flux
and later the scaling between the REF and the NEWs. The local sky bias
was found in annuli of 2-6 times the FWHM and removed prior to
PSF-fitting with the DAOPHOT routines (\citealt{stetson87}; ported to
IDL by \citealt{landsman89}). We calibrated the magnitude scale on the
REF, and thereby all the data, via the USNO-B1.0 R2 magnitudes for the
REFSTARS. The convolved REF was subtracted from the scaled NEWs to
generate subtracted frames (SUBs), and the PSF template measured on
the NEWs were then fit on the SUBs to find the SN flux. Although we
only considered a $30' \times 30'$ region centered on the SN, some
variation in the PSF was noted across the subframes and we decided to
use a spatially varying kernel as described by \citet{alard2000},
which did render improved results. Limiting magnitudes were calculated
for a 1 FWHM radius aperture from the noise measured in the
sky-annuli. The photometric observations and calculated magnitudes are
detailed in Table \ref{SN2006bp_photobs}.

As we first registered the individual images in the NEWs to the REF,
after the image subtraction we obtained multiple, semi-independent
measurements of the SN position on the REF. To determine the
equatorial position of the SN, we matched the REFSTARS to the
USNO-B1.0 and solved for the transformation coefficients. SN~2006bp is
located at $\alpha=11^h53^m55\fs74$,
$\delta=+52\arcdeg21\arcmin09\farcs7$ (J2000.0), with an estimated
error of $0\farcs2$ in each coordinate. We found the location of the
SN on a pre-explosion SDSS r-band frame by directly matching it to the
REF and then transforming the SN position to the SDSS frame. As shown
in Figure \ref{finder}, the SN was found coincident with a red knot in
an outer spiral arm of NGC 3953, and just to the north of a possible
\ion{H}{2} region.

\subsection{ROTSE-III Light Curve}\label{lc}

\begin{figure}
\epsfig{file=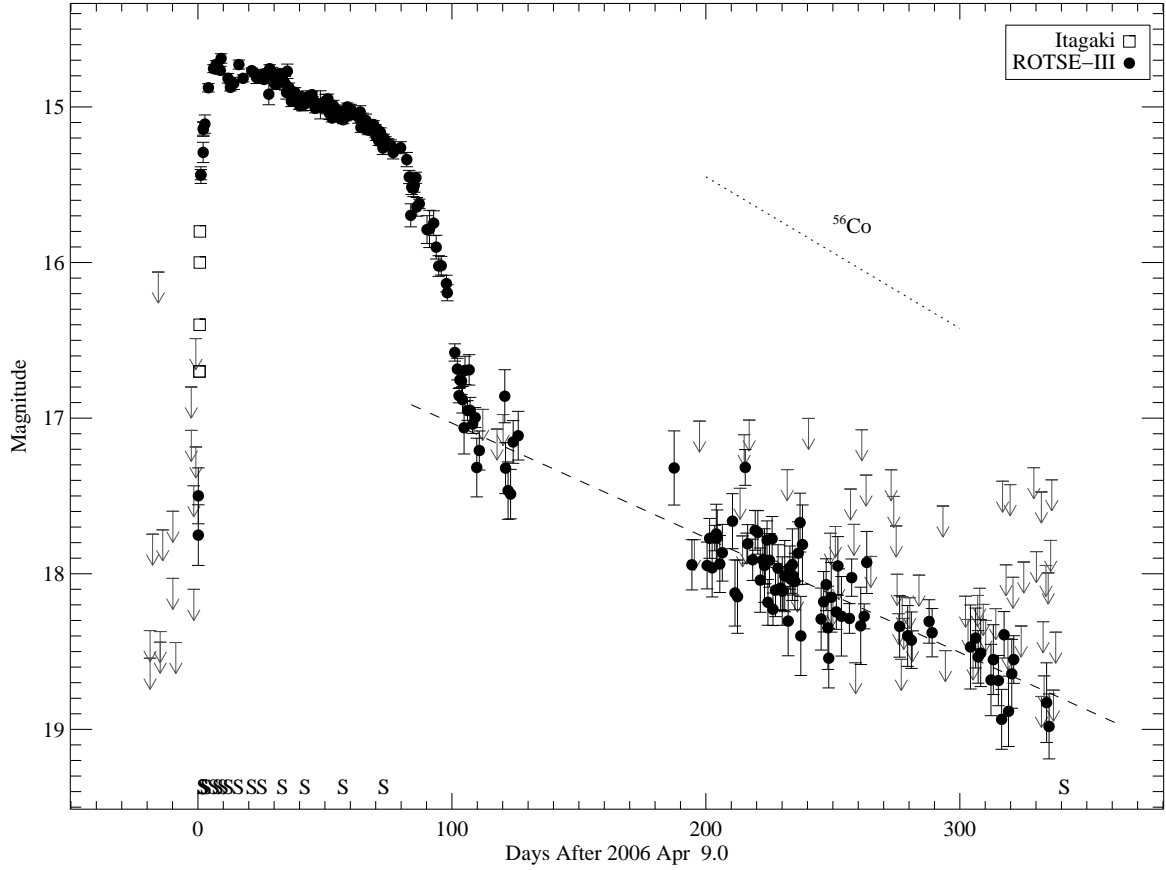,angle=90,width=\columnwidth}
\caption{ROTSE-III unfiltered light curve of SN 2006bp. Detections are
plotted with filled circles and arrows mark the 4-$\sigma$ upper
limits (for clarity, only limits fainter that 17 are shown after day
150). Open squares mark the observations of Mr. Itagaki
\citep{itagaki2006}. The dashed line shows the best fit linear decay
of $0.0073 \pm 0.0004$~mag~day$^{-1}$ between days $+121$ and $+335$,
and the dotted line shows the decay rate for an arbitrary mass of
$^{56}$Co. The S's along the abscissa mark the spectral observation
epochs.}
\label{SN2006bp_lc}
\end{figure}

The unfiltered light curve of SN 2006bp is shown in Figure
\ref{SN2006bp_lc}. ROTSE-III data are plotted as filled circles, and
open squares are used to mark the discovery photometry from
Mr. Itagaki. Because this object is located in the overlap of three of
our search fields, we typically have at least 2 observations per night
covering the position of SN 2006bp in the weeks leading up to its
discovery. We can therefore rule out any bright optical emission prior
to our first detection as might be expected from the ejecta colliding
with a dense stellar wind, energy released from the shock breakout, or
a thermal wave precursor (see \citealt{nadyozhin2003} and references
therein).

\begin{figure}
\plotone{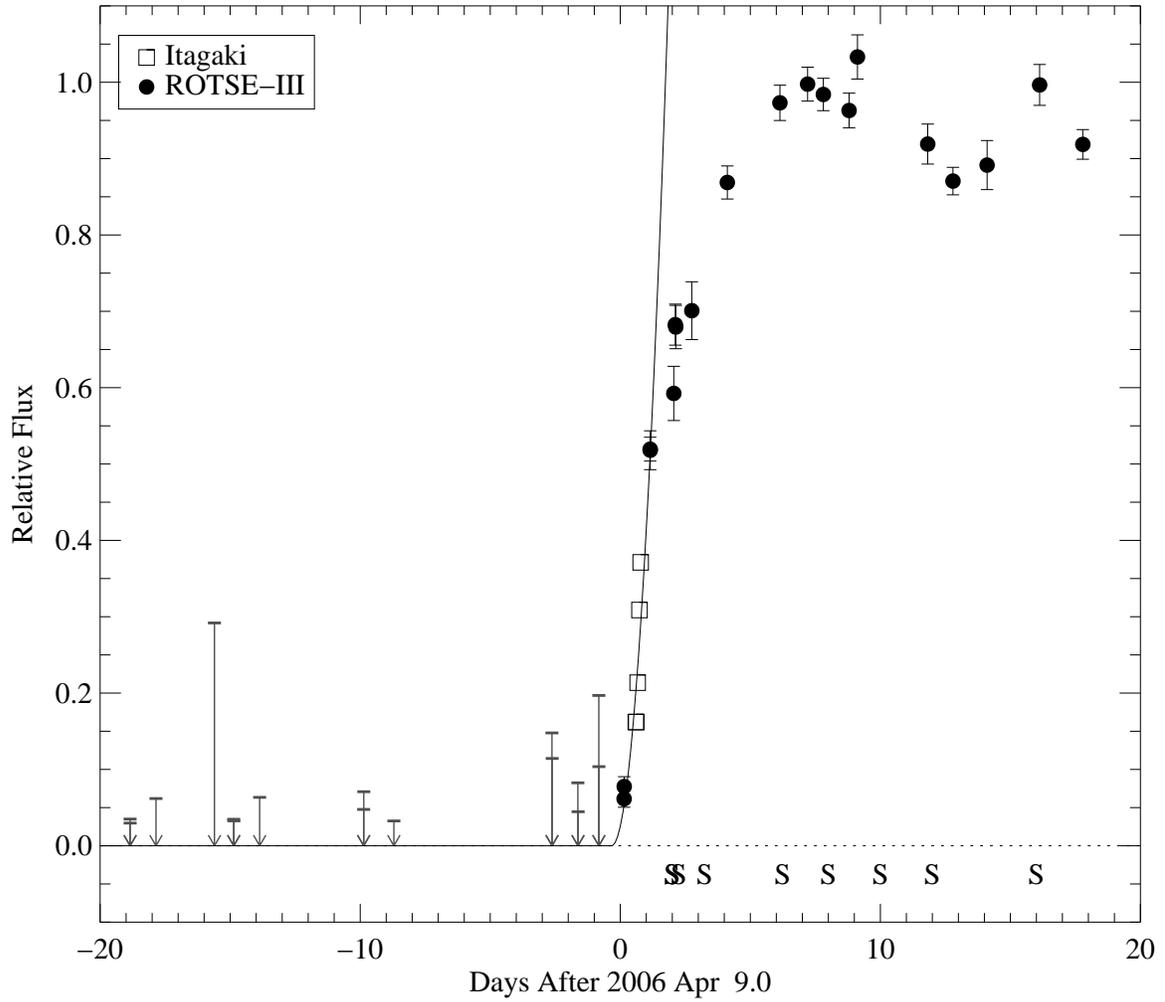}
\caption{Early-time light curve of SN 2006bp in flux space relative to
the 2006 April 16.5 maximum. Arrows mark the 4-$\sigma$ flux limits for
the null-detections (see text). The curve shows the simplistic $F
\propto (t-t_0)^2$ model fit, where $t_0=$April 8.7.}
\label{early_lc}
\end{figure}

The PSF-fit flux and 4-$\sigma$ upper limits of SN~2006bp between 2006
March 20 and April 29 are shown relative to the peak flux in Figure
\ref{early_lc}. Given the initial rapid increase in flux, it is safe
to assume SN~2006bp was detected soon after the explosion shock
emerged from the progenitor envelope; however, detailed modeling tuned
to the observations is required to accurately pinpoint the moment of
core collapse and the subsequent shock breakout. Here we estimate the
date of shock breakout from the early ROTSE-III flux upper limits and
the detections on the first two nights. We modeled the early light
curve as

\begin{equation}
  F(t) = \left\{
  \begin{array}{ll}
    0            &  \rm{for}\; t \le t_0 \\
    A(t-t_0)^2   &  \rm{otherwise}
  \end{array}
  \right.
\end{equation}

\noindent and performed a least squares fit to determine the date of
shock breakout, $t_0$. The fit is plotted as the curve in Figure
\ref{early_lc} (note we fit the model to the flux measured in 1 FWHM
circular apertures instead of the PSF-fit fluxes which are not
statistically accurate for low significance or non-detections,
although the PSF-fit detections are shown in the figure). Under the
simplistic assumption of a quadratic rise, the best fit date for the
shock breakout is 2006 April 8.7. The figure shows an inflection point
in the data around April 10, and extending the fitting range beyond
this date results in an unacceptably large $\chi^2$, which indicates
the model assumptions are not valid over longer periods.  Lacking the
true functional form for the shock breakout light curve in our
unfiltered band pass, we have chosen to reference all phases in this
work relative to 2006 April 9.0 for convenience. It is important also
to note that the actual core collapse can occur 1-3 days before the
shock breakout depending on the progenitor radius.

The ROTSE-III light curve of SN~2006bp rises 3.0 mag from the first
detection on April 9.1 to reach a peak unfiltered magnitude of 14.7
mag on April 16.5 (day +7.5; estimated from a 3$^{\rm rd}$ order
polynomial fit to the data). The light curve then fades by about 0.2
mag before possibly rising to a second maximum on April 25 (day +16),
although the coverage flanking this point is incomplete. There is
otherwise little ($< 0.1$ mag) evolution between days +12 and +33,
after which there is a slight decline of 0.016~mag~day$^{-1}$ until
day +42. The average decay rate from day +38 to day +61 is
0.006~mag~day$^{-1}$, and this rate then increases to
0.013~mag~day$^{-1}$ through day +82. The plateau phase finally comes
to an end around day +80, and SN 2006bp fades by about 2 magnitudes
before entering an exponential decay phase on day +110.

Rejecting a single outlier, the late-time ROTSE-III light curve shows
an average linear decline of $0.0073 \pm 0.0004$~mag~day$^{-1}$
between days +121 and +335, which is significantly slower than the
0.0098~mag~day$^{-1}$ decline expected from the decay of $^{56}$Co
into $^{56}$Fe (e-folding time $\tau_{\rm Co}=111 d$). The late-time
quasi-bolometric and $V$-band light curves of most CCSNe have been
shown to follow the decay of $^{56}$Co, although the $U$- and $B$-band
flux drops more slowly and the $R$- and $I$-band light curves show
departures as well (see \citealt{hamuy1990} for example). The
ROTSE-III data are unfiltered, but the peak response is in the $V$-
and $R$- bands (see Table \ref{rotse_response}). The higher energy
bands contribute little to the measured signal as shown by the low
flux observed at early times when the spectral peak lies in the UV
(see \S\ref{bolo}). It is unclear how the low level $U$- and $B$-band
fluxes at late times (see \S\ref{spec} for the spectra) could dictate
the light curve behavior\footnote{For comparison, the Type II-P
SN~2004et faded by 0.0064~mag~day$^{-1}$ in the $B$-band in the
nebular phase while the $VRI$ rates were all $>
0.01$~mag~day$^{-1}$\citep{sahu2006}.}. It is possible that our
reference template, which was constructed by removing the SN light
from post-explosion data, may not accurately reflect the background
signal around the SN; however, in such case we would expect the light
curve to flatten out or drop quickly once the discrepancy is larger
than the SN flux, but the data show a smooth linear decline. It is
therefore worth considering the possibility of an additional source of
energy production sustaining the luminosity, such as an ejecta-wind
interaction as proposed by \citet{chugai1991} to explain the slower
than expected decline in the H$\alpha$ line flux for three SNe~II at
late times.

\section{Spectroscopy}\label{spec}

We observed SN~2006bp with the Low Resolution Spectrograph (LRS;
\citealt{hill1998}) on the Hobby-Eberly Telescope (HET;
\citealt{ramsey1998}), located at the McDonald Observatory, 13 nights
between 2006 April 11 and June 21, and again on 2007 March 14 and
16. The HET can rotate freely in the azimuthal plane, but it is fixed
at 55 degrees in elevation (airmass $\sim 1.22$), and it uses an
Arecibo-type tracker to follow targets as they rise through the
observable altitude window in the east, or set through it in the
west. The HET has an effective aperture of 9.2-m, although the pupil
can vary along a given track (e.g. the effective mirror size drops
toward the track limits). We used the g1 grism on LRS, which covers
4050~\AA\ to 11000~\AA\ with a resolving power of $R \sim 300$ in a
2.0$''$ slit. To avoid order overlap, we observed SN~2006bp each night
with both the GG385 and the OG590 order sorting filters. The
spectroscopic observing log is detailed in Table
\ref{SN2006bp_specobs}.

Dome flats and arc lamp calibration data were obtained on all nights
with each observing setup; however, standard stars were only observed
on 7 of the first 13 nights. Drawing on past experience with the LRS
that shows the instrument to be quite stable from night to night, we
decided to combine all the available calibration data and use them to
uniformly process the target spectra from each night. We constructed
master flats from all the data available and compared this to the
nightly flats to verify the stability. For the GG385 data, which we
trim to 4100-9200~\AA, the flats vary by less than 1\% from night to
night. We noticed significant variation in the OG590 setup near the
cutoff frequency, however, which seems to correlate with the ambient
(filter) temperature. We trimmed the OG590 data to the 6270-10820~\AA\
range, which keeps the nightly variations below 1\%.

The target spectra were trimmed, bias subtracted, and flat corrected
in the usual manner with IRAF scripts\footnote{IRAF is distributed by
the National Optical Astronomy Observatories, which are operated by
the Association of Universities for Research in Astronomy, Inc., under
cooperative agreement with the National Science Foundation.}. We used
the routine LACosmic \citep{vandokkum2001} to remove particle events,
and then combined the frames from each night in each filter prior to
extraction with APALL. We used spectral observations of the standard
stars BD+17 4708, BD+26 2606, and Wolf 1346 to derive a relative
spectrophotometric correction for the data. After grey shifting, the
different standards taken on different nights agree out to 10000~\AA\
to typically better than 0.05 mag.

We removed strong telluric absorption features using the averaged
BD+26 2606 spectrum as reference. The atmospheric conditions varied
significantly over the course of observations, and the relative
strengths of different telluric features changed from night to
night, so we separately removed features predominantly caused by
\ion{O}{1} before correcting for the H$_{2}$O lines. We note the
H$_{2}$O line profiles change somewhat as the individual transitions
move in and out of saturation, and removal of the strong H$_{2}$O
absorption band between 8900~\AA\ and 9800~\AA\ is imperfect.

\begin{figure}
\epsscale{0.75}
\plotone{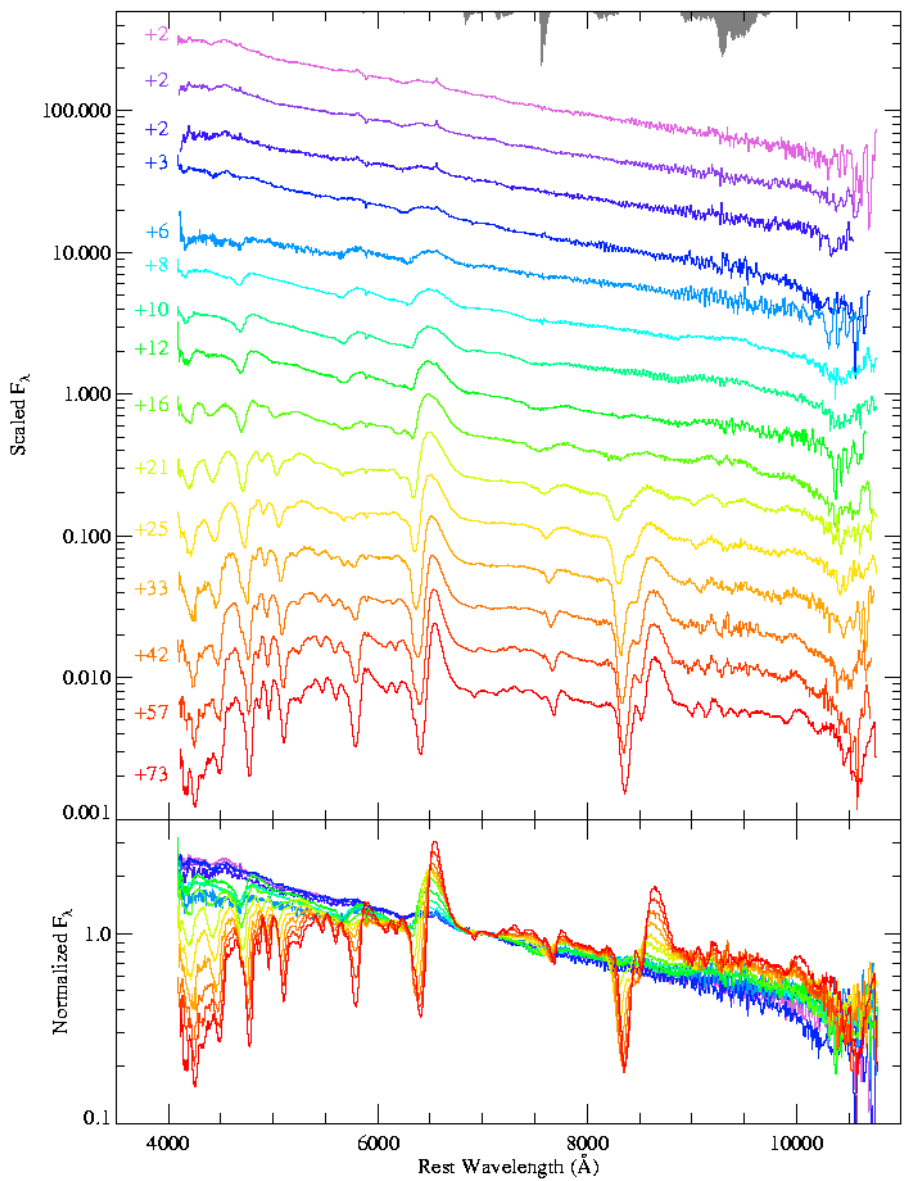}
\caption{Spectral Evolution of SN 2006bp through day +73. The top
portion of the figure shows each spectrum scaled by an arbitrary
factor for clarity and arranged with the earliest spectra on top and
the latest on the bottom. Portions of the spectra suffering from low
signal to noise have been smoothed for display purposes (thin
lines). Phases relative to 2006 April 9.0, a convenient epoch close to
our estimated date of shock breakout, and rounded to the nearest day
are given to the left of each spectrum. For the observations on the
first night (day $+2$), the spectra from both the east and west tracks
are plotted and the combined data are shown in between. Note the $+6$
day spectra were obtained under non-spectroscopic conditions, which
may explain the abrupt departure in the continuum slope as compared to
days $+3$ and $+8$. The bottom of the figure shows each spectrum
normalized around 7000~\AA\ to emphasize the evolution of the line
features and the general reddening with time.}
\label{SN2006bp_all}
\end{figure}

\begin{figure}
\epsscale{1.0}
\plotone{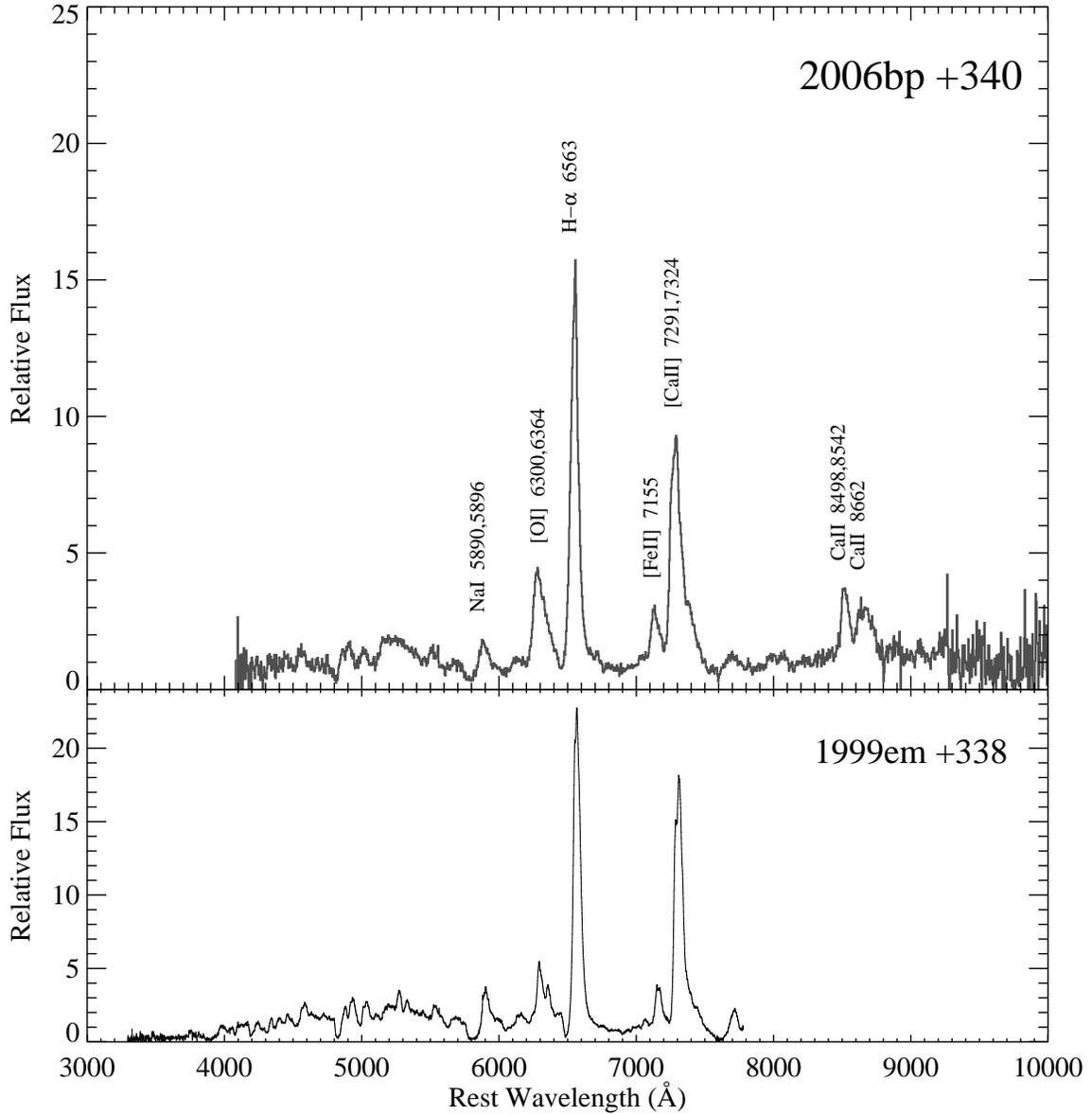}
\caption{HET/LRS spectra of SN~2006bp in the nebular phase
(top). Plotted is the combined spectrum from the 2007 March 14 (OG590)
and March 16 (GG385) data. For reference, a spectrum of SN~1999em taken
around 333 days after discovery ($\sim 338$ days after shock breakout;
\citealt{leonard2002}) is shown below and a few of the strongest
emission lines are labeled.}
\label{SN2006bp_late}
\end{figure}

\begin{figure}
\epsscale{0.5}
\plotone{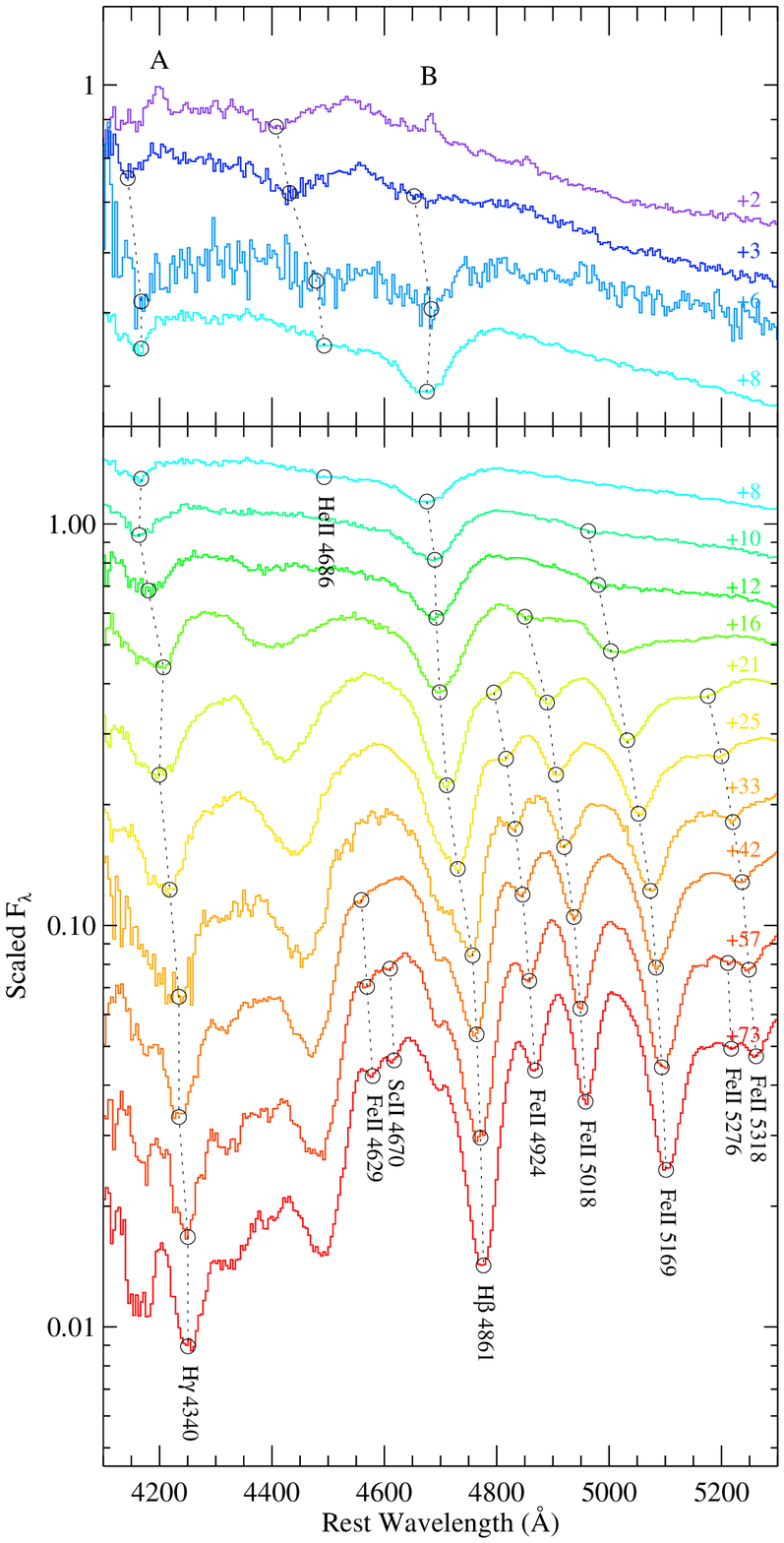}
\caption{Spectral Evolution of SN 2006bp between 4100 and
5300~\AA. The spectra have been arbitrarily scaled for clarity. The
top plot shows the combined $+2$ day spectra along with days $+3$,
$+6$, and $+8$ on an expanded vertical scale to emphasize the subtle
line features. Narrow emission features in the $+2$ day spectra are
marked with the letters ``A'' and ``B.'' We identify B with restframe
\ion{He}{2} $\lambda$4686. Feature A may in part be due to \ion{He}{2}
$\lambda$4200, although it is considerably broader than the
\ion{He}{2} $\lambda$4686 line and may be blended with another
feature. By day $+3$ feature B has disappeared and feature A is not
clearly detected. The $+8$ day spectrum is repeated in the lower plot
along with additional HET spectra through day $+73$ (phases are
labeled along the right of the figure). The FT smoothed minima for key
features found as described in \S\ref{line_evol} are circled, and
dotted lines connect a given minimum through all epochs to guide the
eye.}
\label{SN2006bp_range1}
\end{figure}

From the \ion{H}{1} maps of \citet{vs2001}, we find a recession
velocity of 1,280~km~s$^{-1}$ at the location of SN 2006bp, which we
adopt as the rest frame velocity. To combine the two channels, we
first scaled the OG590 data by a constant and re-sampled it to match
the GG385 spectra in the overlap region. The nightly data are
de-redshifted into the heliocentric rest frame and combined with a
weighted average to produce the final spectra. To ensure a smooth
transition, the weights drop exponentially toward the end of each
component spectrum in the overlap region. For example, the weight of
the GG385 spectra drops from 50\% at 7700~\AA\ to zero at 9200~\AA. On
the first night of observations, we obtained spectra in both the east
and west tracks. We separately analyze each spectrum as well as the
combined spectrum. The fully processed spectra are plotted in Figures
\ref{SN2006bp_all} and \ref{SN2006bp_late}. Figures
\ref{SN2006bp_range1}-\ref{SN2006bp_range4b} show the spectral
evolution on expanded scales to emphasize the evolution of individual
spectral lines.

\subsection{Early Spectra}\label{early_spec}

The HET data from 2006 April 11 and April 12 (about 2 and 3 days after
shock breakout) not only rank among the earliest of supernova
observations, but to our knowledge they represent the earliest spectra
of any normal supernova to date (i.e. excluding the peculiar
SN~1987A). On top of the largely featureless $+2$ day spectra are
narrow emission lines corresponding to restframe \ion{He}{2}
$\lambda$4200, \ion{He}{2} $\lambda$4686, H$\beta$, \ion{C}{4}
$\lambda\lambda$5805, and H$\alpha$ that are each detected in both the
east and west tracks. Table \ref{narrow_lines} gives the measured
equivelent widths and line fluxes for the combined day +2 data. Note
the \ion{He}{2} $\lambda$4200 line appears broader than the
$\lambda$4686 line, which may indicate a blend. Other intermediate
lines from the Pickering series are not detected. In the day $+3$
spectra, H$\alpha$ is clearly detected, although the broad H$\alpha$
P-Cygni emission has increased such that the line is less pronounced,
and the H$\beta$ line is similarly lost in the broad H$\beta$ P-Cygni
emission, but the \ion{He}{2} and \ion{C}{4} lines have disappeared
despite little change or even a decrease in the local continua. The
narrow H$\alpha$ emission is further detected at later phases but the
high ionization lines are not. One possible explanation for this
behavior is that all the narrow lines originate from circumstellar gas
or within a wind blown off the progenitor, although given the location
of the SN in a spiral arm and the proximity to several possible
\ion{H}{2} regions, it is likely the narrow H$\alpha$ emission
originates in a physically distinct region. The narrow \ion{He}{2} and
\ion{C}{4} emission lines, however, require high temperatures or an
ionizing flux, which implies they are local to the SN explosion. Their
disappearance by day $+3$ suggests either rapid recombination at
densities larger than $\approx 10^9$~particles~cm$^{-3}$, which is
characteristic of extended red supergiant atmospheres, or that the
material has been swept up by the shock (the low velocities exclude
the SN ejecta as a source for these narrow lines).

In addition to the narrow emission lines, the day $+2$ and $+3$
spectra also exhibit a limited number of broad absorption
features. Figure \ref{SN2006bp_early} shows these early spectra
plotted in velocity space relative to H$\alpha$. A P-Cygni profile is
seen in each case extending out to beyond 20,000~km~s$^{-1}$, and the
absorption and emission components both strengthen in the latter
spectra. Near the broad emission peak, there are apparent absorption
or emission features in addition to the narrow rest frame H$\alpha$
line. Most prominent among these is a notch extending from 4,000 to
6,500~km~s$^{-1}$. Heeding the cautionary tale of
\citet{matheson2000}, we checked the high resolution atmospheric
transmission atlas from \citet{hinkle2003} and find that uncorrected
telluric lines are likely responsible.

Perhaps the most distinct feature in the day $+2$ spectra is a P-Cygni
profile around 4450~\AA\ (Figures \ref{SN2006bp_all} and
\ref{SN2006bp_range1}). The feature is also present in the day $+3$
spectra, which is similar to a spectrum of SN 1999gi taken one day
after discovery \citep{leonard2002b}. \citet{leonard2002b} tentatively
associate this feature to high velocity H$\beta$, although they
caution that no corresponding high velocity H$\alpha$ profile is
observed. In our early spectra of SN~2006bp we similarly do not
identify P-Cygni profiles from any other species in the same velocity
range (see Figure \ref{SN2006bp_early}). If this feature does in fact
arise from H$\beta$, then the absorption wing extends out to about
35,000~km~s$^{-1}$. This P-Cygni profile taken together with the
narrow \ion{He}{2} $\lambda$4686 emission line is rather evocative of
the H$\alpha$ P-Cygni profile and the narrow H$\alpha$ emission line
(Figure \ref{SN2006bp_early}). Noting this similarity, we suggest the
4450~\AA\ P-Cygni profile may alternatively arise from \ion{He}{2}
$\lambda$4686. This would require high ionization in the outer ejecta
layers, which is plausible given the recent passage of the breakout
shock. Following the modeling of SN~1999em by \citet{baron2000},
\citet{leonard2002b} also identify a broad absorption trough around
5450~\AA\ from high velocity \ion{He}{1} $\lambda$5876 in their early
spectra of SN~1999gi. While this absorption also appears in our $+3$
day spectra of SN~2006bp, it is absent in the $+2$ day data (top of
Figure \ref{SN2006bp_range3}; around 24,000~km~s$^{-1}$ relative to
\ion{He}{1} $\lambda$5876 in Figure \ref{SN2006bp_early}). If the
5450~\AA\ absorption trough is caused by \ion{He}{1} $\lambda$5876,
then its appearance on day $+3$ in concert with the fading of the
proposed \ion{He}{2} $\lambda$4686 P-Cygni line would be consistent
with recombination of Helium. \citet{dessart_hillier2005} suggest
photospheric \ion{Ni}{2} lines can simultaneously fit the features
blue of both \ion{He}{1} $\lambda$5876 and H$\beta$ in spectra of
SN~1999em taken around 5 days after shock breakout (see also
\citealt{baron2000}). Detailed modeling is encouraged to
accurately identify these features.

\subsection{Line Evolution}\label{line_evol}

We compared the HET spectra of SN 2006bp to a spectral atlas of the
Type II-P SN 1999em to identify the line features
(\citealt{leonard2002}; their Figure 10 and Table 4). For this work we
have chosen to focus on the strongest features as well as select lines
useful for EPM studies. To determine the wavelength of maximum
absorption for each feature, we first smoothed the spectra with a
Fourier Transform and divided by the local continuum estimated from a
linear extrapolation to the neighboring peaks. We then interpolated
the data into 0.01~\AA\ bins with a spline function and selected the
flux minima. These wavelengths are then converted into expansion
velocities using the rest wavelength (or $gf$ weighted rest wavelength
for blends) and the relativistic Doppler transformation. Derived
values for the 20 lines chosen are reported in Table \ref{vel_table}
and plotted in Figure \ref{SN2006bp_vels}.

Figure \ref{SN2006bp_range1} shows a detail of the spectral evolution
between 4100 and 5300~\AA. The top portion shows the +2 day to +8 day
spectra on an expanded scale, and the +8 to +73 day spectra are
plotted below. Only the combined (east and west tracks) spectrum is
shown for day +2. The FT smoothed minima of key line features are
circled and labeled. We identify the absorption around 4450~\AA\ in
the +2 day data as a signature of \ion{He}{2} $\lambda$4686 P-Cygni as
described above. By day +8 this feature is almost completely lost in
the continuum, which indicates why this may not have been previously
identified in spectra of other SNe~II-P that were less promptly
observed. H$\beta$ is weakly detected in the day +3 and +6 spectra,
but it is blended with the P-Cygni emission from \ion{He}{2}
$\lambda$4686, which dilutes the feature and may shift the minimum to
the red. By day +73 H$\beta$ is the strongest feature in this
range. Note the kink in the blue wing of H$\beta$ that first becomes
distinct in the +33 day spectrum and persists through day
+73. \ion{Fe}{2} lines begin to emerge 2-3 weeks after shock breakout,
and \ion{Sc}{2} $\lambda$4670 is detected beginning around day
+57. H$\gamma$ is likely blended with \ion{Fe}{2}, \ion{Sr}{2}, and
\ion{Sc}{2} lines. Features left unidentified, such as the absorption
around 4500~\AA\ in the later spectra, are mostly due to blended metal
lines including \ion{Fe}{2}, \ion{Sc}{2}, \ion{Ba}{2}, and
\ion{Ti}{2}.

The spectral evolution of SN~2006bp between 5300 and 6300~\AA\ is
shown in Figure \ref{SN2006bp_range3}. There is interstellar
\ion{Na}{1} absorption present at all phases, suggesting some degree
of reddening along the line of sight. A linear fit to the equivelent
width of this feature over time indicates a 0.005~\AA~day$^{-1}$
decrease from a starting value of $\approx1.3$~\AA\ on April 9.0. The
earliest spectra show a weak \ion{He}{1} $\lambda$5876 P-Cygni profile
that strengthens through day +10. On day +12 this line shows a clear
double minimum in the absorption trough. The blue component to the
pair strengthens and dominates the other, weakening minimum in the day
+21 spectrum. The redder component then strengthens on day +25,
clearly becomes the stronger of the two starting on day +33, and
subsequently develops into the \ion{Na}{1} D P-Cygni feature that
dominates this wavelength range in the latter phases. We have labeled
this red component as \ion{Na}{1} $\lambda$5892 from day +12 through
day +73, although this feature is likely blended with \ion{He}{1}
$\lambda$5876 throughout, and \ion{He}{1} $\lambda$5876 may even be
the stronger component prior to day +21. We further labeled the
minimum around 5700~\AA\ as \ion{He}{1} $\lambda$5876, although this
requires velocities much higher than the other photospheric
lines. Interestingly, this line does drift to the red with time, but
considerably more slowly than the neighboring \ion{Sc}{2}
$\lambda$5533 and \ion{Fe}{2} $\lambda$5666 lines. We note that the
absorption around 5700~\AA\ is much stronger for SN~2006bp then was
seen for SN~1999em, and which \citet{leonard2002} identify as high
velocity \ion{Na}{1}. Detailed modeling is required to accurately
determine the source of the 5700~\AA\ absorption, be it high velocity
\ion{He}{1} $\lambda$5876 or \ion{Na}{1} $\lambda$5892, \ion{N}{2}
$\lambda$5679, a combination of these, or something else. The unmarked
absorption features around 6100~\AA\ and 6200~\AA\ are due to
\ion{Fe}{2}, \ion{Sc}{2}, and \ion{Ba}{2} blends.

Figure \ref{SN2006bp_range2} shows the spectral evolution of SN 2006bp
between 6000 and 7000~\AA\ including the H$\alpha$ P-Cygni
profile. The day +2 to day +6 spectra are shown on an expanded scale
at the top of the figure. H$\alpha$ is detected over all epochs. There
is a kink along the blue wing of the absorption around 6350~\AA\ that
clearly becomes visible in the last 3 epochs. There is also a notch
around 6280~\AA\ in the +57 and +73 day spectra which may be traced
back to day +16.  \citet{leonard2002} identify a similar feature in SN
1999em as high velocity H$\alpha$ and \citet{chugai2007} argue that
such absorption features observed in the spectra of SNe 1999em and
2004dj result from eject-wind interactions. \ion{Si}{2} $\lambda$6533
absorption is detected between days +12 and +25, and there is formally
a minimum found near this line in the +10 day spectra, but given the
large jump in velocities between the +10 and +12 day minima, the
identification with \ion{Si}{2} is dubious (see however
\S\ref{conclusions}).

Fringing becomes a problem in the HET spectra to the red of about
8000~\AA\ as shown in Figure \ref{SN2006bp_range4b}. No lines are
detected in the 6800 to 10300~\AA\ range until day +8. Likely due to
their limited wavelength coverage, the only feature listed by
\citet{leonard2002} to the red of H$\alpha$ is the \ion{Ca}{2} IR
triplet. This feature is weakly detected in the +16 day spectra, and
it quickly grows in strength to dominate this wavelength range. In the
later phases shown in the figure, there is a distinct double minimum
as the two bluer lines blend into \ion{Ca}{2} $\lambda$8520, but the
redder component, \ion{Ca}{2} $\lambda$8662, remains distinct. This
complicated blend and pronounced emission make for a difficult
continuum estimation, and the \ion{Ca}{2} $\lambda$8662 line is
particularly impacted. The components are not clearly split in the +16
day spectrum, and we have associated the absorption only with
\ion{Ca}{2} $\lambda$8520 in the figure, although the 8662~\AA\ line
is likely blended in as well and pulls the minimum to the red as a
result. We identify the P-Cygni feature that becomes clearly visible
around 7700~\AA\ in the later phases in our HET spectra of SN 2006bp
as the \ion{O}{1} 7773~\AA\ triplet. An alternative identification is
a blend of \ion{Fe}{2} lines at 7690, 7705, and 7732~\AA\
\citep{branch2004}. With either identification or a combination of
both, this feature can be seen as early as day +8. Note in the +10 day
spectrum the absorption profile is asymmetric with the red wing
extending much further from the minimum than the blue wing. On day
+21, there is an extended {\it blue} wing next to the deepening
absorption core. A P-Cygni profile from Paschen $\delta$ also becomes
apparent in the latter spectra. Several notches are seen in the +73
day spectra and these may also be present in the +42 and +57 day
spectra, although fringing and night sky line subtraction errors
distort these features. \citet{fassia1998} identify three of these
lines in the Type II-P SN~1995V at a similar phase as \ion{C}{1}
$\lambda$9088, \ion{Sc}{2} $\lambda$9236, and \ion{C}{1}
$\lambda$9405. A possibly distinct set of P-Cygni features are seen
between +21 and +33 days.

As shown in Figure \ref{SN2006bp_all}, the HET spectra extend to
almost 10800~\AA\ in the restframe, although the signal to noise per
pixel can be low beyond $\sim$10300~\AA. There is, however, a clear
absorption dip seen in the +8 and +10 day spectra with a minimum at
about 10400~\AA. Infrared spectra of SN~1999em also show such a dip at
a similar phase, and \citet{hamuy2001} identify this feature as
\ion{He}{1} $\lambda$10830. At later phases (about 25 and 35 days
after breakout), \citet{hamuy2001} find multiple absorption and
emission features in the 10000 to 11000~\AA\ range with possible
contributions from \ion{Sr}{2} $\lambda$10327, \ion{Fe}{2}
$\lambda$10547, and \ion{C}{1} $\lambda$10695 (see also
\citealt{fassia1998}). Another possible contributor to the absorption
in this range is high velocity \ion{He}{1} $\lambda$10830
\citep{chugai2007}.

\section{Bolometric Light Curve Approximation}\label{bolo}

Our ROTSE-III photometry are invaluable in determining the shock
breakout date, but at early times the unfiltered optical response is
insensitive to the bulk of the flux, which is emitted at higher
frequencies. In this section we attempt to reconstruct the
quasi-bolometric light curve of SN~2006bp using the X-ray,
Ultra-violet, and optical observations of \citet{immler2007} and our
HET spectroscopy.

We begin by interpolating the {\it Swift} UVOT measurements in the
UVW2, UVM2, UVW1, U, B, and V filters presented by \citet{immler2007}
onto a grid spaced regularly in time. We fit 4$^{\rm{th}}$ order
polynomials to the light curves over the first 2-5 weeks and
1$^{\rm{st}}$ to 2$^{\rm{nd}}$ order fits to the data after days +10
to +20 as the coverage allows, and smoothly averaged the fits in the
overlap. We then convert the measured magnitudes into flux densities
using the conversion factors available from the {\it Swift}
calibration
database\footnote{\url{http://swift.gsfc.nasa.gov/docs/heasarc/caldb/swift/docs/uvot/}}.
We calibrated our HET spectra by normalizing the flux densities to the
{\it Swift} $V$-band measurements. We then proceeded to interpolate
the flux densities in each wavelength bin to the same regular time
grid using a similar method of smoothly connected polynomial fits
across the early (3$^{\rm{rd}}$ order) and late (2$^{\rm{nd}}$ order)
phases. The HET spectra obtained under non-spectroscopic conditions on
day +6 were excluded from the fits.  For the {\it Swift} XRT and XMM
data \citep{immler2007}, we simply perform a linear fit to the three
epochs of declining flux.

To approximate the bolometric light curve, we then integrate the
1600-5500~\AA\ flux densities from the {\it Swift} photometry and our
HET flux densities between 5500-10000~\AA, and then add in the X-ray
flux, for each time bin. The results are plotted in Figure
\ref{bolo_lc}. As shown by \citet{immler2007}, the spectral energy
distribution (SED) at early-times rises from the optical to the UV
(their Figure 5). Because the 0.2-10 keV X-ray flux is much lower, we
can deduce the peak lies between the X-ray and UVW2 bands on day 1,
but lacking constraint on the SED in this range, we have not attempted
to interpolate the flux densities over the coverage gap. The
early-time bolometric fluxes are therefore somewhat underestimated.
As constructed, the bolometric light curve reaches its peak about 3
days after shock breakout and slowly declines up to day +25 before
leveling off on the plateau.

\section{Discussion and Conclusions}\label{conclusions}

We have presented the ROTSE-III unfiltered light curve and HET optical
spectroscopy for the Type II-P SN~2006bp. Because of our frequent
coverage of the explosion site in the days leading up to shock
breakout, and with the rapid increase in flux observed on 2006 April 9
and 10, we can set the most stringent observational limits ever for
the shock breakout date of a SNe~II-P. A simple quadratic fit to the
early limits and the detections on the first two nights gives an
estimate of 2006 April 8.7 as the date of shock breakout, which is just
11 hours prior to our first detection, although proper modeling of the
light curve with a realistic density structure of the outer layers is
required to accurately determine the breakout epoch. The flux does
increase rapidly between the first two epochs, but the rate of
brightening decreases over the next few nights before reaching a
maximum of 14.7 mag around April 16.5. No distinct breakout peak is
seen in ROTSE-III's unfiltered band pass, and our limits rule out any
such bright optical emission in the 3 days before our first detection.

{\it Swift} began observations of SN~2006bp on 2006 April 10.54 in the
optical, UV, and X-ray bands~\citep{immler2007}. Over the first few
nights, the U, B, and V band curves all rise slightly, and no breakout
peak or rapid decay from such is seen, consistent with the ROTSE-III
data. This behavior excludes the models from \citet{falk_arnett1977}
with extended, low density shells exterior to the progenitor envelope
(i.e. their models B, E, and F), as such a configuration leads to a
pronounced, 10-15 day peak in visible wavelengths. Further the UVW2,
UVM2, and UVW1 light curves decline slowly and do not show any signs
of a distinct, 1-2 day breakout peak as expected for models A and D
from \citet{falk_arnett1977}. The bolometric light curve approximated
in \S\ref{bolo} shows a 3 day rise followed by a 22 day decline to a
plateau phase. Although the early peak flux may be somewhat
underestimated, this asymmetric peak rises just a factor of $\sim 2$
above the plateau. Any sharp jump in luminosity associated with the
arrival of the shock at the progenitor's photosphere must have lasted
less than one day.

The HET spectra are unprecedented not only for the early response, but
also for the broad (4100-10800~\AA), uniform coverage across all
epochs. We employed the same instrument and setups for all
observations from early through late times in contrast to the hodge
podge of observations typically collected. The spectra obtained on
April 11, just two days after the derived date of shock breakout,
represent the earliest spectroscopic observations of a Type II-P, took
place at a similar phase to the first observations of SN~1993J
\citep{lewis1994}, and are the earliest of any supernova other than
SN~1987A. The most intriguing features in these early spectra are
narrow emission lines corresponding to the high ionization species
\ion{He}{2} $\lambda$4686, \ion{C}{4} $\lambda\lambda$5805, and
possibly \ion{He}{2} $\lambda$4200 as well. These features all appear
in our first two epochs of data separated by $\sim$5 hours, but they
have vanished by the following night. The coincidence of these narrow
lines in both space and time with SN~2006bp, their rapid evolution,
and the high temperatures or bright ionizing flux required to produce
them all suggest the emitting region is located close to the explosion
site, although their low velocities and narrow profiles imply this
region is positioned beyond the hydrodynamic front two days after
breakout.  The sudden disappearance of the narrow \ion{He}{2} and
\ion{C}{4} emission lines by day +3 requires either rapid
recombination or disruption of the emission region through direct
interaction with the SN ejecta. Given the lack of narrow emission
lines from lower ionization species on subsequent nights
(e.g. \ion{He}{1} or \ion{C}{3}), we favor the latter
interpretation. We can then determine the distance from the
progenitor's photosphere at the moment of shock breakout to the
emitting region using the observed velocities from H$\alpha$ P-Cygni
in the day +2 spectra (15,000~km~s$^{-1}$ measured from the absorption
minimum, or 21,000~km~s$^{-1}$ as measured from the blue limit of the
absorption). The material responsible for the narrow emission lines in
the day +2 data must lie beyond $3 \times 10^{14}$ cm, but within $5
\times 10^{14}$ cm of the progenitor's surface. If this material were
cast off the progenitor during a stellar eruption and traveled at
10~km~s$^{-1}$, then this occurred just 10-20 years prior to the
explosion.

After the narrow emission lines, the most pronounced feature in the
April 11 spectra is the P-Cygni profile at around 4450~\AA. Other
studies of SNe~II-P with early spectra have noted absorption in this
range and associated it with high velocity H$\beta$, but our earlier
data clearly show such an interpretation requires significant material
at velocities up to 35,000~km~s$^{-1}$. A mechanism to suppress
H$\alpha$ at similar velocities whilst allowing an H$\alpha$ P-Cygni
profile to form at slower velocities would further be required. We
find the identification of this P-Cygni profile with \ion{He}{2}
$\lambda$4686 to be more compelling, and modeling to directly test
this possibility is encouraged. Only weak \ion{He}{1} $\lambda$5876
absorption at velocities consistent with the H$\alpha$ P-Cygni
absorption is found in the day +2 spectra, but by day +3 a distinct
absorption profile appears to the blue of \ion{He}{1} $\lambda$5876 as
well. We hypothesize that this second minimum may form as \ion{He}{2}
in the outer ejecta recombines to \ion{He}{1}, although much of the
corresponding velocity range for the presumed high velocity
\ion{He}{1} $\lambda$5876 absorption is higher than the \ion{He}{2}
$\lambda$4686 velocity limit observed in the day +2 spectrum, so the
observed absorption regions cannot map directly from one line to
the other. This scenario then predicts \ion{He}{2} $\lambda$4686
absorption at earlier phases in the appropriate velocity range.
Contributions from photospheric metal lines such as \ion{N}{2} should
be considered as well \citep{baron2000,dessart_hillier2005,baron2007}.

\begin{figure}
\epsscale{0.5}
\plotone{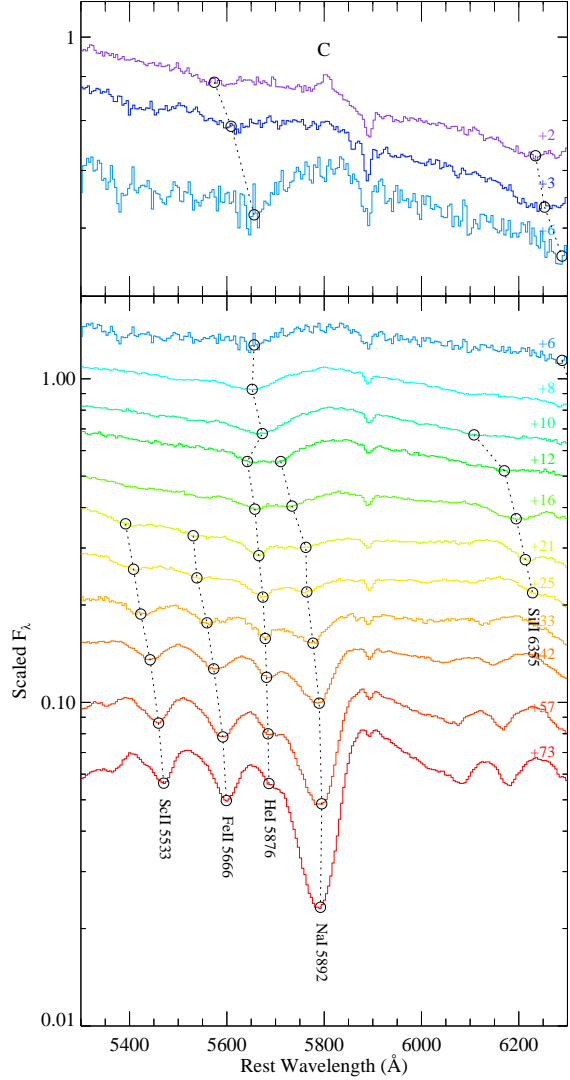}
\caption{Same as figure \ref{SN2006bp_range1} except covering the 5300
to 6300~\AA\ range and only showing the first 3 nights in the top
plot. The letter ``C'' marks a narrow emission feature in the day $+2$
spectra which we identify as the restframe \ion{C}{4}
$\lambda\lambda$5805 doublet. This line is not seen in the day $+3$
spectrum nor in subsequent observations. Over days $+8$ to $+12$ an
absorption line appears around 5700~\AA, which we have labeled as
\ion{He}{1} $\lambda$5876; however, the photospheric \ion{He}{1}
$\lambda$5876 is likely blended with the \ion{Na}{1} $\lambda$5892
feature. The 5700~\AA\ absorption may actually be due to high velocity
\ion{He}{1} or \ion{Na}{1}, or to photospheric \ion{N}{2} (see
text). The narrow absorption at 5892~\AA\ is due to restframe
\ion{Na}{1} D.}
\label{SN2006bp_range3}
\end{figure}

\begin{figure}
\epsscale{0.5}
\plotone{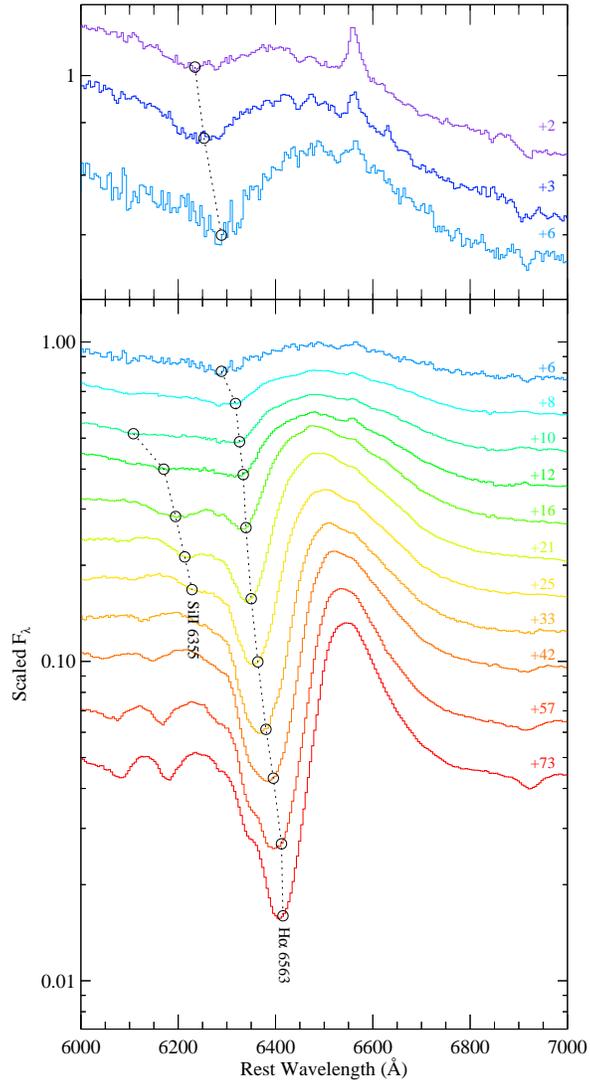}
\caption{Same as figure \ref{SN2006bp_range3} except covering the 6000
to 7000~\AA\ range. Note the narrow H$\alpha$ emission line seen
clearly in the $+2$ day spectrum and which persists atop the broad
H$\alpha$ P-Cygni emission peak into the later phases. The absorption
notch at 6450~\AA\ seen in days +2 and +3 is likely due to uncorrected
telluric absorption.}
\label{SN2006bp_range2}
\end{figure}

\begin{figure}
\epsscale{0.5}
\plotone{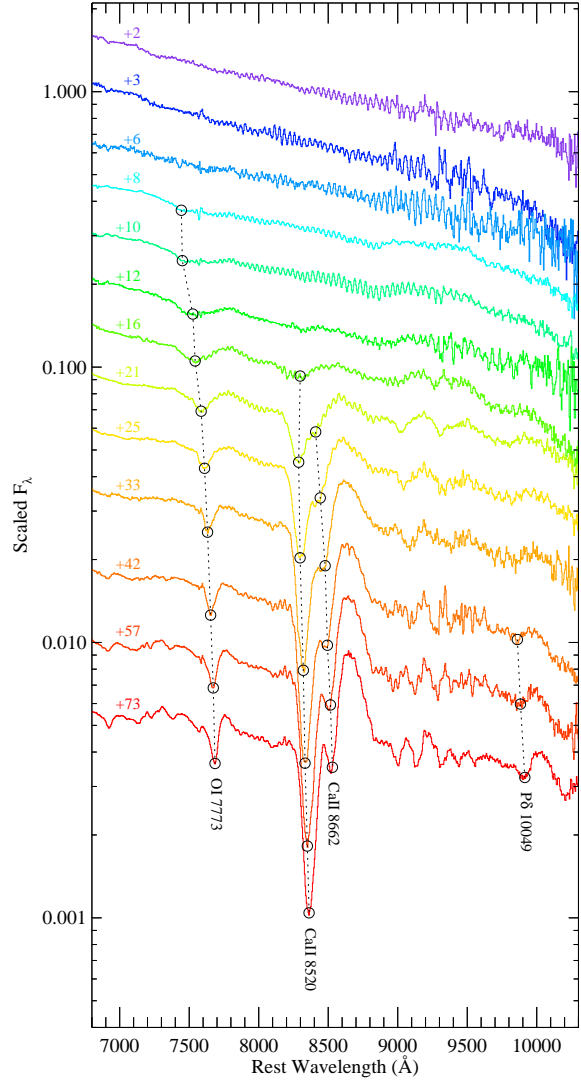}
\caption{Same as figure \ref{SN2006bp_range1} except for the 6800 to
10300~\AA\ range and with all spectra plotted on the same scale (see
figure \ref{SN2006bp_all} for coverage above 10300~\AA). P-Cygni
profiles from P$\delta$ $\lambda$10049 can be seen clearly in day
$+73$ spectra and earlier.}
\label{SN2006bp_range4b}
\end{figure}

\begin{figure}
\epsscale{0.75} \plotone{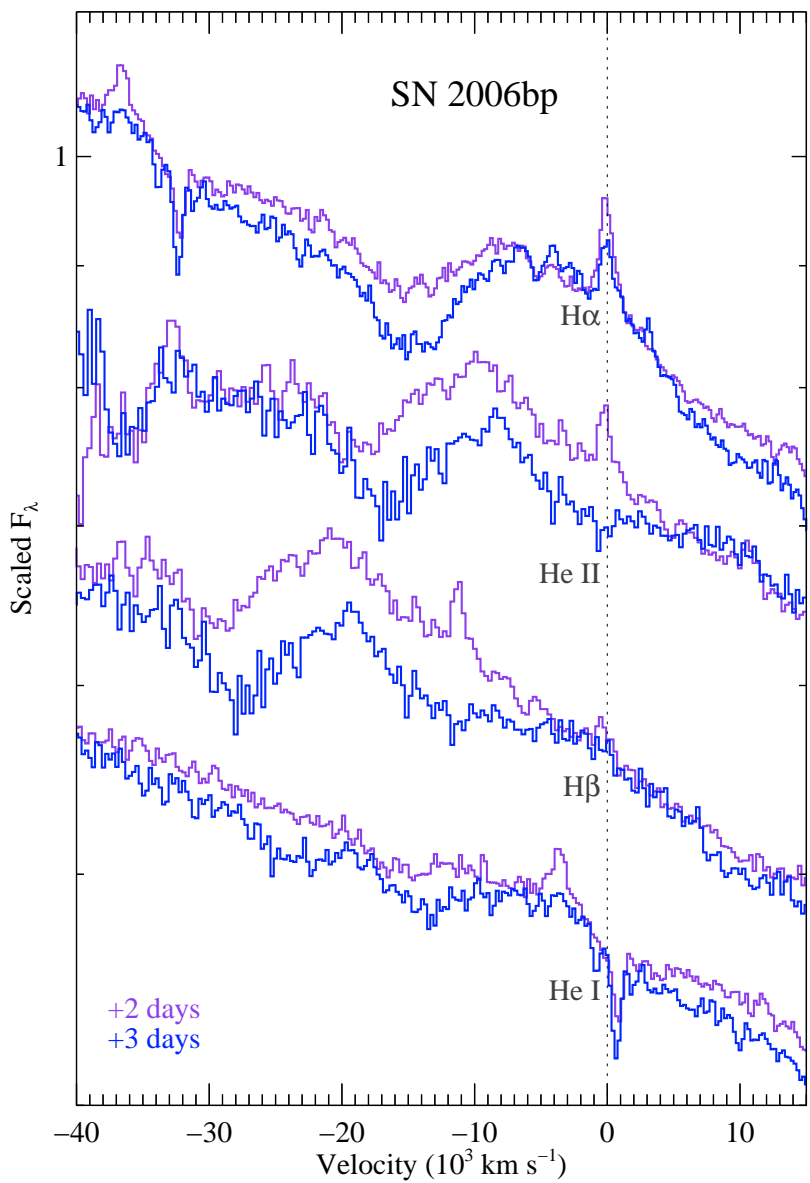}
\caption{Day $+2$ (combined) and day $+3$ spectra of SN~2006bp plotted
in velocity space for (top to bottom) H$\alpha$ $\lambda$6563,
\ion{He}{2} $\lambda$4686, H$\beta$ $\lambda$4861, and \ion{He}{1}
$\lambda$5876 in the adopted 1,280~km~s$^{-1}$ restframe. Notice the
narrow emission features appear biased to the blue, and it is
ambiguous if this shift is caused by an incorrect restframe velocity
for the SN, or of these lines emanate from gas moving relative to the
SN frame. The narrow absorption line seen to the red of rest
\ion{He}{1} $\lambda$5876 is \ion{Na}{1} D in the NGC 3953 rest
frame. The narrow emission feature in the day +2 spectrum blueshifted
by $\sim 3500$~km~s$^{-1}$ relative to \ion{He}{1} $\lambda$5876 is
\ion{C}{4} $\lambda\lambda$5805.}
\label{SN2006bp_early}
\end{figure}

\begin{figure}
\epsscale{1.0}
\plotone{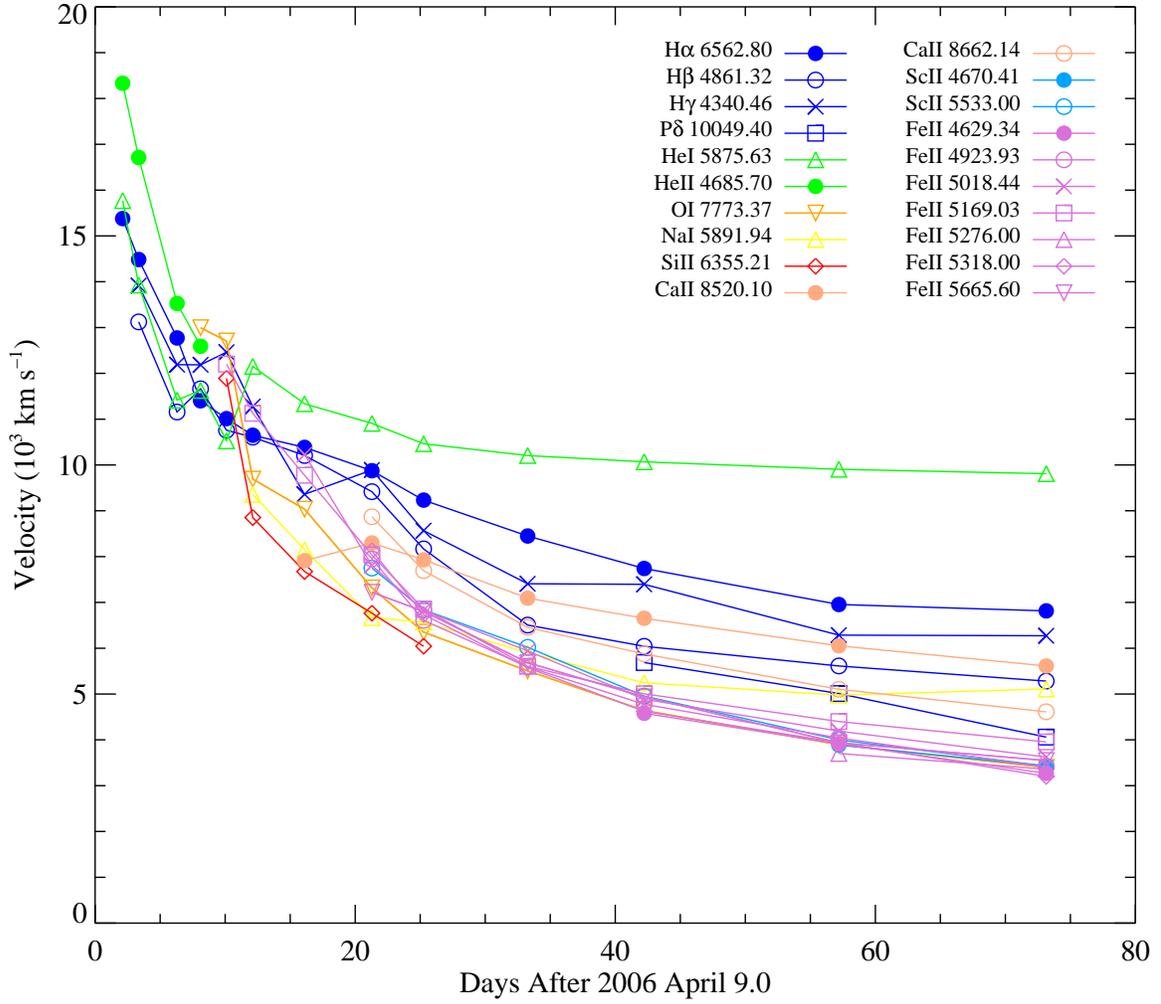}
\caption{Velocities derived from the FT smoothed minima of selected
line features. Symbols mark the measurements and the lines are intended
only to guide the eye.}
\label{SN2006bp_vels}
\end{figure}

\begin{figure}
\epsscale{1.0}
\plotone{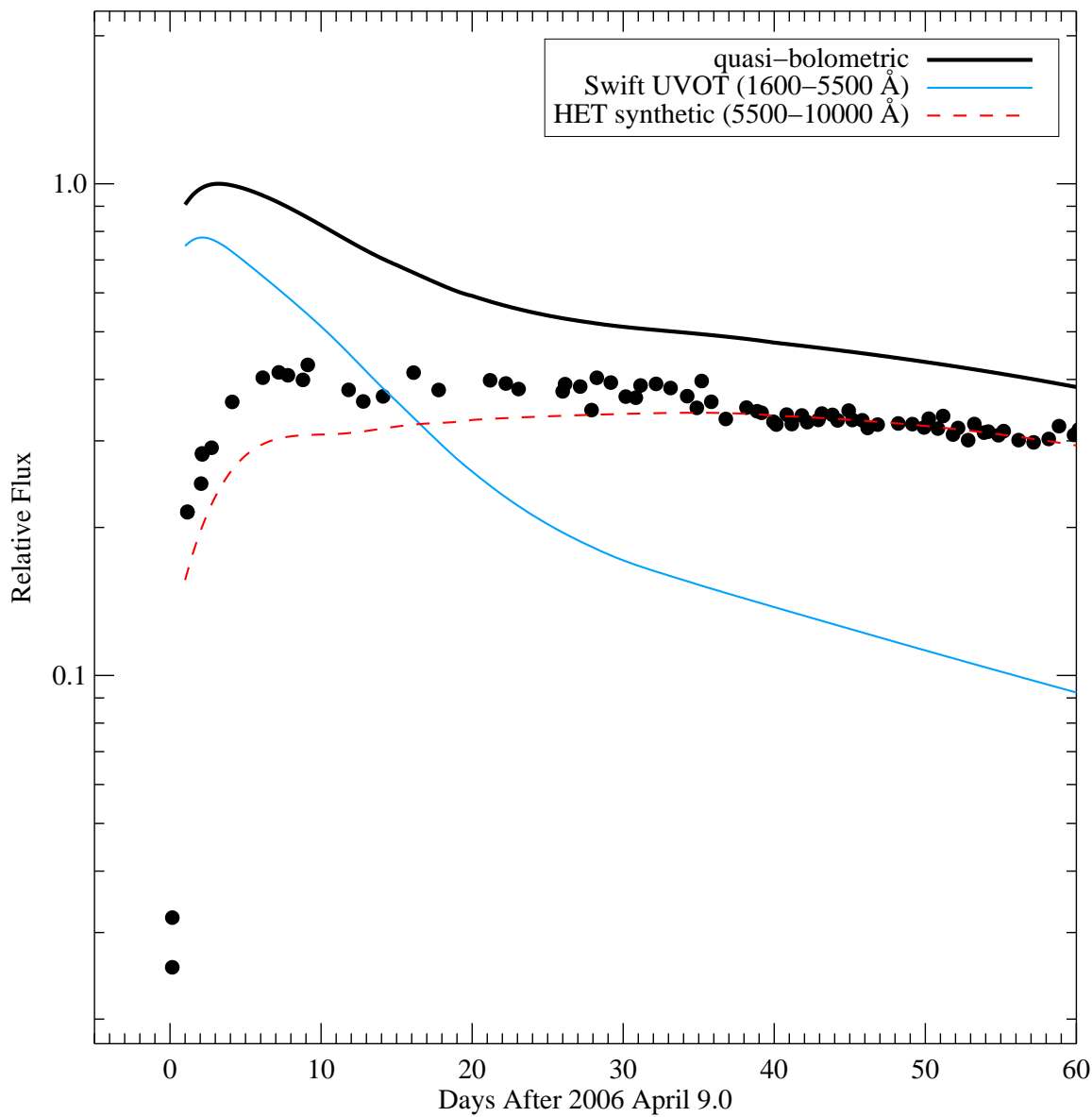}
\caption{Approximate quasi-bolometric light curve (thick black line)
constructed from the X-ray, UV, and optical observations of
\citet{immler2007} and our HET spectra as described in the text. The
0.2-10~keV X-ray contribution included is negligable and the light
curve falls below the plotted range. Note the gap in coverage between
the X-ray and UV bands; at early times the SED peaks in this range,
but we have not attempted to interpolate the flux densities over this
gap, so the early bolometric flux is likely underestimated. The dots
mark the ROTSE-III observations arbitrarily scaled to match the
synthetic HET flux after day +40.}
\label{bolo_lc}
\end{figure}

In the later phases, enhanced absorption appears in the blue wings of
the H$\beta$ (by at least day +33) and \ion{Na}{1} $\lambda$5892 (day
+12) P-Cygni absorption troughs, and in the case of H$\alpha$ there
are two such notches (day +16 and day +42). A possible explanation for
these features are unidentified metal blends, although for the
absorption to the blue of the H$\alpha$ minimum and for the absorption
blue of \ion{Na}{1} $\lambda$5892 (labeled \ion{He}{1} $\lambda$5876
in Figure \ref{SN2006bp_range3}), the weak absorption minima do not
shift to longer wavelengths with time in a manner consistent with the
photospheric evolution. A second possibility is that these lines
originate in higher velocity layers above the photosphere. We find
anecdotal evidence for such a line forming region when considering 1)
the three of these features that appear at about 10,000~km~s$^{-1}$
relative to H$\beta$, \ion{He}{1} $\lambda$5876, and H$\alpha$, and 2)
the \ion{Fe}{2} $\lambda$5018, \ion{Fe}{2} $\lambda$5169, \ion{Si}{2}
$\lambda$6533, and \ion{O}{1} $\lambda$7773 lines that all initially
appear at substantially higher velocities than would be predicted from
backward extrapolation of later phase measurements (see Figure
\ref{SN2006bp_vels}). More intriguing still, the velocities of these
lines upon entrance (days +8 to +10) match the egress velocity of the
proposed \ion{He}{2} $\lambda$4686 line (day +8). SN~2006bp may very
well have had a line forming region exterior to the photosphere,
perhaps formed from the interaction of the progenitor wind as
suggested by \citet{chugai2007}. The initial emergence of certain
species in these high velocity layers could then imply this region
cools faster than the photospheric layers, perhaps signaling a source
of heating such as outward mixing of $^{56}$Ni in the lower line
forming region \citep{fassia1998}. In this rough picture, the ongoing
ejecta/wind interaction could supply additional energy at late times
and thus help explain the slow, $0.0073 \pm 0.0004$~mag~day$^{-1}$
decline derived from the ROTSE-III photometry in the nebular phase.

Finally, we would like to comment on the confusion of the terms
``shock breakout'' and ``explosion,'' which appears to occur quite
frequently in the literature. For Type II supernovae, the explosion
date should be referenced to the moment at which the core reverses its
collapse. The shock breakout, on the other hand, is the first
electromagnetic signal of the explosion and should be referenced to
the moment at which the stellar flux begins its rapid ascent to
super-Eddington luminosities and beyond. This epoch will be delayed
relative to the core collapse as the signal must first reach the
surface of the progenitor. Depending on the extent of the envelope,
this delay may be hours (e.g. SN~1987A), or up to a few days for
super-redgiant envelopes ($\sim$5000 $R_\odot$). Of particular
importance are dates derived from the expanding photosphere method,
where an equation analogous to

\begin{equation}\label{epm}
R = R_0 + v(t-t_0)
\end{equation}

\noindent is solved via least squares fits to give $t_0$ and the
radius of the emitting photosphere, $R$. The initial radius, $R_0$, is
typically assumed to be negligible, and in that same vein, $t_0$ is
often referred to as the explosion date; however, it is clear from
equation \ref{epm} that $t_0$ is actually the epoch at which the
photospheric radius begins to grow and properly refers to the shock
breakout date. As electromagnetic studies push toward earlier and
earlier epochs, and with gravitational wave detectors soon to join the
neutrino experiments as complimentary probes into the deaths of stars,
it is important that we exercise diligence in our use of these
separate terms.

\acknowledgments We would like to thank the staff of the Hobby-Eberly
Telescope and McDonald Observatory for their support. We give specific
thanks to S. Rostopchin, J. Caldwell, M. Shetrone, F. Deglman,
S. Odewahn, V. Riley, and E. Terrazas for their observations with the
HET, and to F. Castro, P. Mondol, and M. Sellers for their efforts in
screening potential SN candidates.  This work made use of the SUSPECT
on-line database of SNe spectra
(\url{http://bruford.nhn.ou.edu/$\sim$suspect/index1.html}).  This
research is supported, in part, by NASA grant NAG 5-7937 (PH) and NSF
grants AST0307312 (PH) and AST0406740 (RQ \& JCW).


\begin{deluxetable}{cc}
\tablewidth{0pt}
\tablecaption{ROTSE-III Sensitivity}
\tablehead{
  \colhead{Wavelength (\AA)} &
  \colhead{Efficiency}
}
\startdata
  3000  &    0.00  \\
  3500  &    0.09  \\
  4000  &    0.45  \\
  4500  &    0.67  \\
  5000  &    0.83  \\
  5500  &    0.84  \\
  6000  &    0.84  \\
  6500  &    0.82  \\
  7000  &    0.78  \\
  7500  &    0.71  \\
  8000  &    0.57  \\
  8500  &    0.40  \\
  9000  &    0.28  \\
  9500  &    0.16  \\
 10000  &    0.07  \\
\enddata
\tablecomments{Approximate spectral response for ROTSE-III not
including atmospheric losses. The values have not been empericly
measured, but were instead calculated based on the typical CCD and optical
element efficiencies. Additional losses due to degredation of the
optical elements over time are not taken into account.}
\label{rotse_response}
\end{deluxetable}
\clearpage

\begin{deluxetable}{crrrrrr}
\tablewidth{0pt}
\tablecaption{ROTSE-III Observations of SN 2006bp}
\tabletypesize{\scriptsize}
\tablehead{
  \colhead{Inst.$^{a}$} &
  \colhead{Phase$^{b}$} &
  \colhead{Date} &
  \colhead{Exp. (s)} &
  \colhead{$C_R$} &
  \colhead{$\sigma_{C_R}$} &
  \colhead{Limit$^{c}$}
}
\startdata
 3b &    -18.8 &    2006 Mar 21.152 &    3x20 &  \nodata &  \nodata &    18.37 \\
 3b &    -18.8 &    Mar 21.154 &    4x20 &  \nodata &  \nodata &    18.54 \\
 3b &    -17.9 &    Mar 22.148 &    3x20 &  \nodata &  \nodata &    17.75 \\
 3b &    -15.6 &    Mar 24.396 &    4x20 &  \nodata &  \nodata &    16.06 \\
 3b &    -14.9 &    Mar 25.135 &    4x20 &  \nodata &  \nodata &    18.37 \\
 3b &    -14.9 &    Mar 25.142 &    4x20 &  \nodata &  \nodata &    18.44 \\
 3b &    -13.9 &    Mar 26.134 &    4x20 &  \nodata &  \nodata &    17.72 \\
 3b &     -9.9 &    Mar 30.132 &    3x20 &  \nodata &  \nodata &    18.03 \\
 3b &     -9.9 &    Mar 30.137 &    2x20 &  \nodata &  \nodata &    17.60 \\
 3b &     -8.7 &    Mar 31.295 &    4x20 &  \nodata &  \nodata &    18.44 \\
 3b &     -2.6 &    Apr  6.373 &    4x20 &  \nodata &  \nodata &    16.80 \\
 3b &     -2.6 &    Apr  6.382 &    2x20 &  \nodata &  \nodata &    17.08 \\
 3b &     -1.6 &    Apr  7.370 &    1x20 &  \nodata &  \nodata &    17.43 \\
 3b &     -1.6 &    Apr  7.379 &    4x20 &  \nodata &  \nodata &    18.10 \\
 3b &     -0.8 &    Apr  8.176 &    1x20 &  \nodata &  \nodata &    16.49 \\
 3b &     -0.8 &    Apr  8.177 &    2x20 &  \nodata &  \nodata &    17.19 \\
 3b &      0.1 &    Apr  9.147 &    4x20 &    17.75 &     0.19 &    17.97 \\
 3b &      0.2 &    Apr  9.151 &    4x20 &    17.50 &     0.18 &    17.69 \\
 3b &      1.1 &    Apr 10.149 &    4x20 &    15.44 &     0.03 &    17.65 \\
 3b &      1.2 &    Apr 10.152 &    4x20 &    15.44 &     0.05 &    17.47 \\
 3d &      2.1 &    Apr 11.058 &    5x20 &    15.29 &     0.07 &    17.43 \\
 3b &      2.1 &    Apr 11.115 &    3x20 &    15.14 &     0.04 &    17.72 \\
 3b &      2.1 &    Apr 11.134 &    4x20 &    15.14 &     0.05 &    17.40 \\
 3d &      2.8 &    Apr 11.751 &    4x20 &    15.11 &     0.06 &    16.97 \\
 3b &      4.1 &    Apr 13.116 &    3x20 &    14.88 &     0.03 &    17.65 \\
 3b &      6.1 &    Apr 15.139 &    6x20 &    14.75 &     0.03 &    18.24 \\
 3b &      7.2 &    Apr 16.202 &    3x20 &    14.73 &     0.02 &    17.71 \\
 3d &      7.8 &    Apr 16.808 &    4x60 &    14.74 &     0.02 &    18.50 \\
 3d &      8.8 &    Apr 17.798 &    3x60 &    14.76 &     0.03 &    17.47 \\
 3b &      9.1 &    Apr 18.120 &    3x20 &    14.69 &     0.03 &    18.34 \\
 3d &     11.8 &    Apr 20.822 &    8x60 &    14.82 &     0.03 &    19.14 \\
 3d &     12.8 &    Apr 21.792 &    5x60 &    14.87 &     0.02 &    18.80 \\
 3d &     14.1 &    Apr 23.103 &    7x60 &    14.85 &     0.04 &    18.14 \\
 3b &     16.1 &    Apr 25.124 &    3x20 &    14.73 &     0.03 &    18.51 \\
 3d &     17.8 &    Apr 26.785 &    8x60 &    14.82 &     0.02 &    18.67 \\
 3b &     21.2 &    Apr 30.190 &    9x20 &    14.77 &     0.02 &    19.12 \\
 3b &     22.2 &    May  1.231 &    6x20 &    14.78 &     0.02 &    18.89 \\
 3d &     23.1 &    May  2.074 &    5x60 &    14.81 &     0.04 &    18.74 \\
 3d &     26.0 &    May  4.994 &    6x60 &    14.82 &     0.02 &    18.67 \\
 3b &     26.2 &    May  5.152 &    6x20 &    14.79 &     0.03 &    18.59 \\
 3b &     27.2 &    May  6.153 &    6x20 &    14.80 &     0.02 &    18.46 \\
 3d &     27.9 &    May  6.905 &    1x60 &    14.92 &     0.07 &    16.44 \\
 3b &     28.3 &    May  7.253 &    3x20 &    14.75 &     0.03 &    17.77 \\
 3b &     29.2 &    May  8.181 &    3x20 &    14.78 &     0.03 &    17.20 \\
 3b &     30.2 &    May  9.170 &    6x20 &    14.85 &     0.04 &    18.29 \\
 3d &     30.8 &    May  9.837 &    8x20 &    14.86 &     0.03 &    17.85 \\
 3b &     31.2 &    May 10.150 &    5x20 &    14.79 &     0.03 &    17.85 \\
 3b &     32.2 &    May 11.176 &    5x20 &    14.79 &     0.03 &    17.65 \\
 3b &     33.1 &    May 12.135 &    3x20 &    14.81 &     0.03 &    17.65 \\
 3b &     34.2 &    May 13.238 &    3x20 &    14.85 &     0.04 &    17.53 \\
 3d &     34.9 &    May 13.874 &    8x20 &    14.91 &     0.04 &    17.92 \\
 3b &     35.2 &    May 14.195 &    2x20 &    14.77 &     0.05 &    16.66 \\
 3d &     35.8 &    May 14.833 &   12x20 &    14.88 &     0.03 &    18.30 \\
 3d &     36.8 &    May 15.793 &    4x60 &    14.96 &     0.03 &    18.78 \\
 3b &     38.2 &    May 17.160 &    6x20 &    14.91 &     0.03 &    18.87 \\
 3d &     38.9 &    May 17.869 &    8x60 &    14.92 &     0.02 &    18.79 \\
 3b &     39.2 &    May 18.160 &    6x20 &    14.93 &     0.03 &    19.07 \\
 3d &     40.0 &    May 18.950 &    4x60 &    14.98 &     0.04 &    18.28 \\
 3b &     40.1 &    May 19.140 &    3x20 &    14.99 &     0.02 &    18.48 \\
 3d &     40.8 &    May 19.823 &   11x60 &    14.94 &     0.03 &    18.96 \\
 3b &     41.2 &    May 20.162 &    6x20 &    14.99 &     0.04 &    18.98 \\
 3d &     41.8 &    May 20.827 &   12x60 &    14.95 &     0.02 &    18.85 \\
 3b &     42.2 &    May 21.184 &    3x20 &    14.98 &     0.04 &    18.74 \\
 3d &     42.9 &    May 21.921 &    7x60 &    14.97 &     0.02 &    18.91 \\
 3b &     43.2 &    May 22.152 &    4x20 &    14.94 &     0.04 &    18.56 \\
 3d &     43.9 &    May 22.852 &    7x60 &    14.94 &     0.02 &    18.82 \\
 3b &     44.2 &    May 23.194 &    5x20 &    14.97 &     0.02 &    18.73 \\
 3d &     44.9 &    May 23.927 &    8x60 &    14.92 &     0.02 &    18.62 \\
 3b &     45.2 &    May 24.174 &    5x20 &    14.97 &     0.03 &    18.85 \\
 3d &     45.8 &    May 24.830 &   12x60 &    14.97 &     0.02 &    19.02 \\
 3b &     46.2 &    May 25.185 &    3x20 &    15.01 &     0.03 &    18.61 \\
 3d &     46.8 &    May 25.845 &    9x60 &    14.99 &     0.03 &    18.84 \\
 3b &     48.2 &    May 27.209 &    2x20 &    14.99 &     0.09 &    16.19 \\
 3b &     49.1 &    May 28.145 &    3x20 &    14.99 &     0.04 &    18.25 \\
 3d &     49.9 &    May 28.920 &   12x60 &    15.01 &     0.03 &    19.17 \\
 3b &     50.2 &    May 29.229 &    5x20 &    14.96 &     0.04 &    18.61 \\
 3d &     50.8 &    May 29.835 &   12x60 &    15.01 &     0.03 &    19.02 \\
 3b &     51.2 &    May 30.178 &    4x20 &    14.95 &     0.03 &    18.42 \\
 3d &     51.8 &    May 30.836 &   11x60 &    15.04 &     0.04 &    19.00 \\
 3b &     52.2 &    May 31.168 &    4x20 &    15.01 &     0.04 &    18.56 \\
 3d &     52.8 &    May 31.836 &   12x60 &    15.07 &     0.02 &    19.04 \\
 3b &     53.2 &    Jun  1.247 &    3x20 &    14.99 &     0.03 &    18.15 \\
 3d &     53.9 &    Jun  1.890 &    6x60 &    15.03 &     0.02 &    18.64 \\
 3b &     54.2 &    Jun  2.169 &    6x20 &    15.03 &     0.04 &    18.60 \\
 3d &     54.8 &    Jun  2.841 &   11x60 &    15.05 &     0.05 &    18.74 \\
 3b &     55.2 &    Jun  3.188 &    6x20 &    15.03 &     0.03 &    18.57 \\
 3b &     56.2 &    Jun  4.178 &    4x20 &    15.07 &     0.03 &    18.40 \\
 3b &     57.2 &    Jun  5.170 &    6x20 &    15.08 &     0.03 &    18.46 \\
 3b &     58.2 &    Jun  6.170 &    6x20 &    15.07 &     0.02 &    18.38 \\
 3d &     58.9 &    Jun  6.852 &    9x20 &    15.00 &     0.03 &    18.14 \\
 3d &     59.8 &    Jun  7.847 &   10x20 &    15.04 &     0.03 &    18.06 \\
 3b &     60.2 &    Jun  8.170 &    4x20 &    15.02 &     0.03 &    17.96 \\
 3d &     60.8 &    Jun  8.831 &   10x20 &    15.05 &     0.03 &    17.88 \\
 3b &     61.2 &    Jun  9.166 &    5x20 &    15.05 &     0.04 &    18.05 \\
 3b &     63.2 &    Jun 11.171 &    3x20 &    15.06 &     0.04 &    17.56 \\
 3d &     63.9 &    Jun 11.886 &    8x20 &    15.03 &     0.04 &    17.22 \\
 3b &     64.2 &    Jun 12.152 &    2x20 &    15.13 &     0.03 &    17.66 \\
 3d &     64.8 &    Jun 12.841 &   12x20 &    15.09 &     0.05 &    18.04 \\
 3b &     65.2 &    Jun 13.163 &    4x20 &    15.10 &     0.02 &    18.26 \\
 3b &     66.2 &    Jun 14.152 &    3x20 &    15.09 &     0.04 &    18.45 \\
 3d &     66.8 &    Jun 14.843 &    4x60 &    15.14 &     0.03 &    18.26 \\
 3b &     67.2 &    Jun 15.178 &    5x20 &    15.15 &     0.03 &    18.63 \\
 3b &     68.2 &    Jun 16.174 &    6x20 &    15.14 &     0.04 &    18.79 \\
 3d &     68.8 &    Jun 16.831 &    5x60 &    15.15 &     0.02 &    18.45 \\
 3b &     69.2 &    Jun 17.165 &    4x20 &    15.11 &     0.02 &    18.27 \\
 3d &     69.9 &    Jun 17.902 &    4x60 &    15.15 &     0.04 &    18.28 \\
 3b &     70.2 &    Jun 18.165 &    4x20 &    15.19 &     0.03 &    18.61 \\
 3d &     70.8 &    Jun 18.818 &    5x60 &    15.15 &     0.02 &    18.40 \\
 3b &     71.2 &    Jun 19.175 &    6x20 &    15.22 &     0.03 &    18.66 \\
 3d &     71.8 &    Jun 19.850 &    8x60 &    15.16 &     0.03 &    18.68 \\
 3b &     72.2 &    Jun 20.182 &    3x20 &    15.22 &     0.03 &    18.54 \\
 3d &     72.8 &    Jun 20.844 &   12x60 &    15.26 &     0.04 &    19.01 \\
 3b &     73.2 &    Jun 21.176 &    6x20 &    15.20 &     0.03 &    18.83 \\
 3d &     73.8 &    Jun 21.841 &   11x60 &    15.22 &     0.03 &    18.85 \\
 3d &     74.9 &    Jun 22.873 &    8x60 &    15.23 &     0.03 &    18.61 \\
 3d &     75.9 &    Jun 23.893 &    4x60 &    15.25 &     0.03 &    18.23 \\
 3d &     76.9 &    Jun 24.858 &    8x60 &    15.29 &     0.04 &    18.64 \\
 3d &     79.8 &    Jun 27.802 &    4x60 &    15.26 &     0.04 &    18.26 \\
 3b &     82.2 &    Jun 30.230 &    3x20 &    15.34 &     0.05 &    18.31 \\
 3b &     83.2 &    Jul  1.198 &    3x20 &    15.45 &     0.04 &    17.92 \\
 3d &     83.8 &    Jul  1.802 &    4x60 &    15.70 &     0.07 &    16.93 \\
 3b &     84.2 &    Jul  2.181 &    6x20 &    15.52 &     0.05 &    18.46 \\
 3d &     84.8 &    Jul  2.836 &    4x60 &    15.52 &     0.05 &    17.41 \\
 3b &     85.2 &    Jul  3.176 &    6x20 &    15.50 &     0.04 &    18.32 \\
 3d &     85.8 &    Jul  3.823 &    8x60 &    15.46 &     0.04 &    18.18 \\
 3b &     86.2 &    Jul  4.176 &    6x20 &    15.64 &     0.06 &    18.31 \\
 3b &     87.2 &    Jul  5.165 &    4x20 &    15.62 &     0.03 &    18.00 \\
 3b &     90.2 &    Jul  8.197 &    3x20 &    15.79 &     0.09 &    16.97 \\
 3b &     91.2 &    Jul  9.170 &    5x20 &    15.78 &     0.12 &    16.85 \\
 3d &     92.8 &    Jul 10.820 &    8x20 &    15.75 &     0.08 &    17.41 \\
 3d &     93.8 &    Jul 11.816 &    7x20 &    15.90 &     0.08 &    17.28 \\
 3d &     94.8 &    Jul 12.819 &    8x20 &    16.02 &     0.07 &    17.29 \\
 3d &     95.8 &    Jul 13.798 &    4x60 &    16.02 &     0.06 &    18.20 \\
 3d &     97.8 &    Jul 15.833 &    4x60 &    16.14 &     0.05 &    18.10 \\
 3b &     98.2 &    Jul 16.171 &    3x20 &    16.19 &     0.05 &    17.40 \\
 3b &    101.1 &    Jul 19.146 &    3x20 &    16.58 &     0.06 &    18.15 \\
 3b &    102.2 &    Jul 20.155 &   12x20 &    16.68 &     0.07 &    18.94 \\
 3d &    102.8 &    Jul 20.808 &    3x60 &    16.86 &     0.05 &    18.05 \\
 3b &    103.2 &    Jul 21.155 &   12x20 &    16.75 &     0.05 &    18.92 \\
 3d &    103.8 &    Jul 21.793 &    4x60 &    16.76 &     0.08 &    18.09 \\
 3b &    104.2 &    Jul 22.161 &    8x20 &    16.88 &     0.09 &    18.45 \\
 3d &    104.8 &    Jul 22.792 &    3x60 &    17.06 &     0.17 &    18.07 \\
 3b &    105.1 &    Jul 23.147 &    6x20 &    16.69 &     0.09 &    17.47 \\
 3b &    106.2 &    Jul 24.153 &   12x20 &    16.95 &     0.06 &    18.74 \\
 3d &    106.8 &    Jul 24.791 &    3x60 &    16.69 &     0.10 &    17.72 \\
 3b &    107.2 &    Jul 25.165 &    3x20 &    16.95 &     0.08 &    18.16 \\
 3b &    108.2 &    Jul 26.152 &   12x20 &    17.04 &     0.06 &    18.86 \\
 3b &    109.1 &    Jul 27.150 &    9x20 &    17.00 &     0.06 &    18.79 \\
 3d &    109.8 &    Jul 27.788 &    3x60 &    17.32 &     0.19 &    17.92 \\
 3d &    110.8 &    Jul 28.787 &    4x60 &    17.21 &     0.13 &    18.42 \\
 3b &    112.1 &    Jul 30.146 &    2x20 &  \nodata &  \nodata &    16.94 \\
 3d &    117.8 &    Aug  4.780 &    2x60 &  \nodata &  \nodata &    17.07 \\
 3b &    120.2 &    Aug  7.156 &    2x20 &  \nodata &  \nodata &    16.98 \\
 3d &    120.8 &    Aug  7.816 &   10x20 &    16.86 &     0.17 &    17.43 \\
 3b &    121.1 &    Aug  8.119 &   16x20 &    17.32 &     0.16 &    17.96 \\
 3b &    122.1 &    Aug  9.118 &   10x20 &    17.47 &     0.18 &    17.64 \\
 3b &    123.1 &    Aug 10.117 &   17x20 &    17.49 &     0.16 &    18.12 \\
 3b &    124.1 &    Aug 11.114 &   11x20 &    17.15 &     0.14 &    18.19 \\
 3b &    126.1 &    Aug 13.111 &   12x20 &    17.11 &     0.16 &    17.65 \\
 3b &    187.5 &    Oct 13.505 &   12x20 &    17.32 &     0.24 &    17.33 \\
 3b &    194.5 &    Oct 20.493 &   33x20 &    17.94 &     0.16 &    19.12 \\
 3b &    195.5 &    Oct 21.485 &    9x20 &  \nodata &  \nodata &    17.78 \\
 3b &    197.5 &    Oct 23.504 &    1x20 &  \nodata &  \nodata &    17.02 \\
 3b &    200.5 &    Oct 26.482 &   32x20 &    17.95 &     0.15 &    19.21 \\
 3b &    201.5 &    Oct 27.480 &    7x20 &    17.77 &     0.13 &    18.29 \\
 3b &    202.5 &    Oct 28.477 &   32x20 &    17.96 &     0.19 &    19.22 \\
 3b &    203.5 &    Oct 29.471 &   24x20 &    17.77 &     0.08 &    19.25 \\
 3d &    204.1 &    Oct 30.108 &    4x60 &    17.74 &     0.19 &    18.10 \\
 3b &    204.5 &    Oct 30.472 &   16x20 &    17.77 &     0.18 &    19.09 \\
 3b &    205.5 &    Oct 31.478 &   24x20 &    17.94 &     0.18 &    19.12 \\
 3b &    206.5 &    Nov  1.476 &   24x20 &    17.87 &     0.18 &    19.07 \\
 3b &    210.5 &    Nov  5.464 &   23x20 &    17.66 &     0.18 &    18.39 \\
 3b &    211.5 &    Nov  6.460 &   23x20 &    18.12 &     0.21 &    18.52 \\
 3b &    212.5 &    Nov  7.458 &   24x20 &    18.15 &     0.24 &    18.65 \\
 3b &    213.5 &    Nov  8.469 &    2x20 &  \nodata &  \nodata &    17.45 \\
 3b &    214.5 &    Nov  9.455 &    7x20 &  \nodata &  \nodata &    17.76 \\
 3d &    215.1 &    Nov 10.072 &    4x20 &  \nodata &  \nodata &    17.11 \\
 3b &    215.5 &    Nov 10.501 &    7x20 &    17.32 &     0.11 &    17.97 \\
 3d &    216.1 &    Nov 11.084 &    4x60 &  \nodata &  \nodata &    17.74 \\
 3b &    216.5 &    Nov 11.453 &   18x20 &    17.81 &     0.12 &    18.51 \\
 3d &    217.1 &    Nov 12.066 &    1x60 &  \nodata &  \nodata &    17.01 \\
 3b &    218.4 &    Nov 13.442 &   23x20 &    17.91 &     0.13 &    18.98 \\
 3b &    219.4 &    Nov 14.427 &    7x20 &    17.72 &     0.13 &    18.29 \\
 3b &    220.5 &    Nov 15.457 &    4x20 &    17.73 &     0.14 &    18.08 \\
 3b &    221.4 &    Nov 16.433 &   24x20 &    18.04 &     0.21 &    19.13 \\
 3b &    222.4 &    Nov 17.428 &    9x20 &    17.91 &     0.15 &    18.34 \\
 3d &    223.1 &    Nov 18.078 &    8x60 &    17.95 &     0.24 &    18.61 \\
 3b &    223.4 &    Nov 18.430 &   24x20 &    17.91 &     0.15 &    18.95 \\
 3d &    224.1 &    Nov 19.097 &    4x60 &    17.78 &     0.12 &    18.38 \\
 3b &    224.4 &    Nov 19.428 &   24x20 &    18.18 &     0.15 &    19.08 \\
 3d &    225.1 &    Nov 20.051 &    4x60 &    17.91 &     0.17 &    18.28 \\
 3d &    226.1 &    Nov 21.076 &    4x60 &    17.78 &     0.14 &    18.25 \\
 3b &    226.4 &    Nov 21.414 &   15x20 &    18.23 &     0.10 &    18.83 \\
 3b &    227.4 &    Nov 22.420 &   24x20 &    18.11 &     0.17 &    19.18 \\
 3b &    228.4 &    Nov 23.417 &   24x20 &    17.96 &     0.15 &    19.04 \\
 3b &    229.4 &    Nov 24.413 &   24x20 &    18.09 &     0.14 &    19.15 \\
 3d &    230.1 &    Nov 25.063 &   12x60 &    18.11 &     0.22 &    18.78 \\
 3b &    230.4 &    Nov 25.411 &   21x20 &    18.11 &     0.13 &    19.09 \\
 3d &    231.1 &    Nov 26.056 &    8x60 &    18.01 &     0.23 &    18.65 \\
 3d &    232.0 &    Nov 27.031 &    4x60 &  \nodata &  \nodata &    17.33 \\
 3b &    232.4 &    Nov 27.404 &   23x20 &    18.30 &     0.22 &    19.12 \\
 3d &    233.1 &    Nov 28.090 &    8x60 &    17.96 &     0.14 &    18.53 \\
 3b &    233.4 &    Nov 28.401 &   15x20 &    18.03 &     0.09 &    18.95 \\
 3d &    234.0 &    Nov 29.049 &    8x60 &    17.94 &     0.23 &    18.65 \\
 3b &    234.4 &    Nov 29.399 &   22x20 &    18.01 &     0.12 &    19.04 \\
 3d &    235.0 &    Nov 30.046 &    8x60 &    18.05 &     0.18 &    18.67 \\
 3d &    236.1 &    Dec  1.064 &    3x60 &  \nodata &  \nodata &    18.05 \\
 3b &    236.4 &    Dec  1.402 &    5x20 &    17.87 &     0.20 &    17.93 \\
 3d &    237.1 &    Dec  2.081 &    3x60 &    17.67 &     0.19 &    17.94 \\
 3b &    237.4 &    Dec  2.389 &   24x20 &    18.40 &     0.25 &    18.69 \\
 3d &    238.0 &    Dec  3.034 &    8x20 &    17.81 &     0.25 &    17.83 \\
 3b &    239.4 &    Dec  4.397 &    3x20 &  \nodata &  \nodata &    16.23 \\
 3b &    240.4 &    Dec  5.422 &    2x20 &  \nodata &  \nodata &    17.00 \\
 3b &    245.4 &    Dec 10.378 &    8x60 &    18.29 &     0.20 &    18.61 \\
 3b &    246.4 &    Dec 11.375 &    8x60 &    18.18 &     0.19 &    18.77 \\
 3b &    247.4 &    Dec 12.367 &    7x60 &    18.07 &     0.17 &    18.49 \\
 3d &    248.1 &    Dec 13.065 &    8x60 &    18.35 &     0.27 &    18.48 \\
 3b &    248.4 &    Dec 13.379 &    6x60 &    18.54 &     0.19 &    18.70 \\
 3d &    249.0 &    Dec 14.014 &    6x60 &  \nodata &  \nodata &    17.74 \\
 3b &    249.4 &    Dec 14.358 &    3x60 &    18.15 &     0.22 &    18.37 \\
 3d &    250.0 &    Dec 15.025 &    4x60 &  \nodata &  \nodata &    18.13 \\
 3b &    250.3 &    Dec 15.344 &    3x60 &  \nodata &  \nodata &    16.66 \\
 3d &    251.0 &    Dec 15.988 &    3x60 &  \nodata &  \nodata &    17.70 \\
 3b &    251.4 &    Dec 16.363 &    6x60 &    18.25 &     0.11 &    18.88 \\
 3d &    252.0 &    Dec 16.979 &    4x60 &    17.95 &     0.19 &    18.37 \\
 3b &    252.3 &    Dec 17.328 &    2x60 &  \nodata &  \nodata &    17.97 \\
 3b &    253.4 &    Dec 18.378 &    5x60 &    18.27 &     0.26 &    18.82 \\
 3b &    256.5 &    Dec 21.505 &    6x60 &    18.29 &     0.10 &    19.30 \\
 3d &    257.0 &    Dec 21.963 &    4x60 &  \nodata &  \nodata &    17.46 \\
 3b &    257.4 &    Dec 22.413 &    4x60 &    18.02 &     0.12 &    18.79 \\
 3b &    258.3 &    Dec 23.313 &    1x60 &  \nodata &  \nodata &    17.68 \\
 3d &    259.0 &    Dec 23.980 &    8x60 &  \nodata &  \nodata &    18.57 \\
 3d &    261.0 &    Dec 25.974 &    8x60 &    18.33 &     0.25 &    18.46 \\
 3b &    261.4 &    Dec 26.393 &    2x60 &  \nodata &  \nodata &    17.07 \\
 3b &    262.3 &    Dec 27.334 &    7x60 &    18.27 &     0.08 &    19.31 \\
 3d &    263.1 &    Dec 28.086 &    3x60 &  \nodata &  \nodata &    17.37 \\
 3b &    263.3 &    Dec 28.342 &    1x60 &    17.93 &     0.20 &    18.16 \\
 3d &    265.0 &    Dec 29.961 &    8x20 &  \nodata &  \nodata &    17.89 \\
 3d &    272.9 &    2007 Jan  6.939 &    8x20 &  \nodata &  \nodata &    17.33 \\
 3d &    274.0 &    Jan  7.990 &    9x20 &  \nodata &  \nodata &    17.50 \\
 3d &    274.9 &    Jan  8.934 &    8x20 &  \nodata &  \nodata &    17.69 \\
 3b &    275.3 &    Jan  9.295 &    8x20 &  \nodata &  \nodata &    18.00 \\
 3d &    275.9 &    Jan  9.933 &    8x60 &  \nodata &  \nodata &    18.26 \\
 3b &    276.3 &    Jan 10.291 &    6x60 &    18.34 &     0.20 &    18.44 \\
 3d &    276.9 &    Jan 10.942 &   10x60 &  \nodata &  \nodata &    18.55 \\
 3b &    277.3 &    Jan 11.330 &    2x60 &  \nodata &  \nodata &    18.15 \\
 3d &    277.9 &    Jan 11.939 &    5x60 &  \nodata &  \nodata &    18.29 \\
 3b &    278.4 &    Jan 12.416 &    2x60 &  \nodata &  \nodata &    16.57 \\
 3b &    279.5 &    Jan 13.455 &    4x60 &    18.40 &     0.20 &    18.53 \\
 3b &    280.3 &    Jan 14.273 &    2x60 &  \nodata &  \nodata &    18.15 \\
 3d &    280.9 &    Jan 14.920 &    8x60 &    18.43 &     0.18 &    18.64 \\
 3b &    281.4 &    Jan 15.360 &    2x60 &  \nodata &  \nodata &    18.37 \\
 3d &    282.1 &    Jan 16.148 &    1x60 &  \nodata &  \nodata &    16.55 \\
 3d &    283.9 &    Jan 17.916 &    5x60 &  \nodata &  \nodata &    18.01 \\
 3d &    287.9 &    Jan 21.899 &    8x60 &    18.31 &     0.14 &    18.67 \\
 3d &    289.1 &    Jan 23.077 &    8x60 &    18.38 &     0.16 &    18.75 \\
 3b &    293.3 &    Jan 27.346 &    1x60 &  \nodata &  \nodata &    17.56 \\
 3b &    294.3 &    Jan 28.312 &    8x20 &  \nodata &  \nodata &    18.49 \\
 3b &    295.2 &    Jan 29.240 &    4x20 &  \nodata &  \nodata &    16.68 \\
 3b &    296.3 &    Jan 30.269 &    1x20 &  \nodata &  \nodata &    16.67 \\
 3d &    297.9 &    Jan 31.927 &    7x20 &  \nodata &  \nodata &    16.93 \\
 3b &    302.2 &    Feb  5.222 &    8x20 &  \nodata &  \nodata &    18.14 \\
 3b &    303.2 &    Feb  6.219 &    8x20 &  \nodata &  \nodata &    18.29 \\
 3b &    304.2 &    Feb  7.185 &    2x60 &    18.47 &     0.27 &    18.48 \\
 3b &    305.2 &    Feb  8.215 &    4x40 &  \nodata &  \nodata &    18.48 \\
 3b &    306.2 &    Feb  9.216 &    5x60 &    18.41 &     0.19 &    19.06 \\
 3d &    307.0 &    Feb 10.041 &    3x60 &  \nodata &  \nodata &    18.14 \\
 3b &    307.2 &    Feb 10.211 &    8x60 &    18.53 &     0.17 &    19.18 \\
 3d &    307.8 &    Feb 10.848 &    7x60 &  \nodata &  \nodata &    18.09 \\
 3b &    308.2 &    Feb 11.209 &    7x60 &    18.51 &     0.21 &    19.16 \\
 3b &    309.2 &    Feb 12.215 &    2x60 &  \nodata &  \nodata &    18.20 \\
 3b &    310.2 &    Feb 13.203 &    4x60 &  \nodata &  \nodata &    18.30 \\
 3b &    312.3 &    Feb 15.265 &    4x60 &    18.68 &     0.23 &    18.81 \\
 3b &    313.2 &    Feb 16.195 &    8x60 &    18.55 &     0.22 &    19.12 \\
 3b &    314.2 &    Feb 17.159 &    2x60 &  \nodata &  \nodata &    18.22 \\
 3b &    315.2 &    Feb 18.188 &    8x60 &    18.69 &     0.16 &    19.31 \\
 3b &    316.5 &    Feb 19.454 &    7x60 &    18.94 &     0.19 &    19.25 \\
 3d &    316.8 &    Feb 19.824 &    4x60 &  \nodata &  \nodata &    17.41 \\
 3b &    317.4 &    Feb 20.422 &    5x60 &    18.39 &     0.15 &    19.00 \\
 3b &    318.2 &    Feb 21.197 &    4x60 &  \nodata &  \nodata &    17.94 \\
 3d &    318.9 &    Feb 21.936 &    4x60 &  \nodata &  \nodata &    18.42 \\
 3b &    319.2 &    Feb 22.179 &    8x60 &    18.88 &     0.23 &    19.19 \\
 3d &    319.8 &    Feb 22.791 &    4x60 &  \nodata &  \nodata &    17.43 \\
 3b &    320.5 &    Feb 23.453 &    6x60 &    18.64 &     0.22 &    19.16 \\
 3d &    320.9 &    Feb 23.935 &    3x60 &  \nodata &  \nodata &    18.02 \\
 3b &    321.2 &    Feb 24.211 &    4x60 &    18.55 &     0.15 &    18.66 \\
 3b &    324.2 &    Feb 27.162 &    8x20 &  \nodata &  \nodata &    18.34 \\
 3b &    325.2 &    Feb 28.164 &    7x20 &  \nodata &  \nodata &    17.92 \\
 3b &    326.3 &    Mar  1.326 &    1x20 &  \nodata &  \nodata &    16.04 \\
 3b &    327.1 &    Mar  2.139 &    5x20 &  \nodata &  \nodata &    16.65 \\
 3b &    328.1 &    Mar  3.134 &    3x20 &  \nodata &  \nodata &    16.52 \\
 3b &    329.1 &    Mar  4.138 &    6x20 &  \nodata &  \nodata &    17.32 \\
 3b &    330.1 &    Mar  5.145 &    8x20 &  \nodata &  \nodata &    17.86 \\
 3d &    332.1 &    Mar  7.101 &    8x20 &  \nodata &  \nodata &    17.47 \\
 3b &    332.1 &    Mar  7.131 &    4x60 &  \nodata &  \nodata &    18.79 \\
 3d &    332.8 &    Mar  7.786 &    7x60 &  \nodata &  \nodata &    18.31 \\
 3b &    333.2 &    Mar  8.165 &    3x60 &  \nodata &  \nodata &    18.66 \\
 3d &    333.9 &    Mar  8.911 &    8x20 &  \nodata &  \nodata &    17.95 \\
 3b &    334.2 &    Mar  9.163 &    6x60 &    18.83 &     0.26 &    19.13 \\
 3d &    334.8 &    Mar  9.804 &    5x60 &  \nodata &  \nodata &    17.99 \\
 3b &    335.1 &    Mar 10.134 &    8x60 &    18.98 &     0.21 &    19.24 \\
 3d &    335.8 &    Mar 10.771 &    4x60 &  \nodata &  \nodata &    17.79 \\
 3b &    336.2 &    Mar 11.208 &    2x60 &  \nodata &  \nodata &    17.40 \\
 3d &    336.9 &    Mar 11.854 &    8x60 &  \nodata &  \nodata &    18.75 \\
 3d &    337.8 &    Mar 12.762 &    8x60 &  \nodata &  \nodata &    18.38 \\
\enddata
\tablenotetext{a}{Instrument responsible for the observation,
abbreviated ``3b'' for ROTSE-IIIb in west Texas and ``3d'' for
ROTSE-IIId in Turkey.}
\tablenotetext{b}{Relative to 2006 April 9.0.}
\tablenotetext{c}{4-$\sigma$ limiting magnitude.}
\label{SN2006bp_photobs}
\end{deluxetable}

\clearpage

\begin{deluxetable}{cccccl}
\tablewidth{0pt}
\tablecaption{Observing Log for HET Spectra of SN~2006bp}
\tabletypesize{\scriptsize}
\tablehead{
  \colhead{Date (UT)} &
  \colhead{JD-2400000.5} &
  \colhead{Phase$^{a}$ (day)} &
  \colhead{Exp. (s)} &
  \colhead{Filter} &
  \colhead{Notes$^{b}$}
}
\startdata
2006 Apr 11.11 &   53836.11 &       2 & 2x300 &   GG385  &  S \\
2006 Apr 11.12 &   53836.12 &       2 & 1x500 &   OG590  &  S, cloudy\\
2006 Apr 11.31 &   53836.31 &       2 & 3x360 &   GG385  &  N, cloudy \\
2006 Apr 11.33 &   53836.33 &       2 & 2x550 &   OG590  &  N, cloudy \\
2006 Apr 12.35 &   53837.35 &       3 & 1x600 &   GG385  &  N, cloudy\\
2006 Apr 12.36 &   53837.36 &       3 & 1x900 &   OG590  &  S, cloudy\\
2006 Apr 15.30 &   53840.30 &       6 & 1x600 &   GG385  &  N, cloudy\\
2006 Apr 15.31 &   53840.31 &       6 & 572,900 & OG590  &  N, cloudy\\
2006 Apr 17.10 &   53842.10 &       8 & 2x200 &   GG385  &  S \\
2006 Apr 17.11 &   53842.11 &       8 & 2x450 &   OG590  &  S \\
2006 Apr 19.10 &   53844.10 &      10 & 2x200 &   GG385  &  S \\
2006 Apr 19.11 &   53844.11 &      10 & 2x450 &   OG590  &  S \\
2006 Apr 21.12 &   53846.12 &      12 & 2x200 &   GG385  &  P \\
2006 Apr 21.13 &   53846.13 &      12 & 400,500 & OG590  &  P \\
2006 Apr 25.10 &   53850.10 &      16 & 2x400 &   GG385  & N, cloudy \\
2006 Apr 25.11 &   53850.11 &      16 & 540,421 &   OG590  & S  \\
2006 Apr 30.28 &   53855.28 &      21 & 2x200 &   GG385  & S  \\
2006 Apr 30.29 &   53855.29 &      21 & 2x450 &   OG590  & S  \\
2006 May  4.26 &   53859.26 &      25 & 2x200 &   GG385  & P  \\
2006 May  4.27 &   53859.27 &      25 & 2x450 &   OG590  & P  \\
2006 May 12.26 &   53867.26 &      33 & 2x200 &   GG385  & S  \\
2006 May 12.27 &   53867.27 &      33 & 500,550 &   OG590  & S  \\
2006 May 21.21 &   53876.21 &      42 & 2x400 &   GG385  & S  \\
2006 May 21.23 &   53876.23 &      42 & 2x600 &   OG590  & S  \\
2006 Jun  5.19 &   53891.19 &      57 & 2x200 &   GG385  & S  \\
2006 Jun  5.20 &   53891.20 &      57 & 2x450 &   OG590  & S  \\
2006 Jun 21.14 &   53907.14 &      73 & 2x200 &   GG385  & P  \\
2006 Jun 21.14 &   53907.14 &      73 & 2x450 &   OG590  & P  \\
2007 Mar 14.21 &   54173.21 &     339 & 2x900,875 &   OG590 & S  \\
2007 Mar 16.17 &   54175.17 &     341 & 650,2x600 &   GG385 & P  \\
\enddata
\tablecomments{All observations were taken through a 2$''$ slit oriented along the paralactic angle.}
\tablenotetext{a}{Relative to 2006 April 9.0 and rounded off to the
nearest day.}
\tablenotetext{b}{Sky conditions at the time of
observation (P=photometric, S=spectroscopic, N=non-spectroscopic)}
\label{SN2006bp_specobs}
\end{deluxetable}

\begin{deluxetable}{rcc}
\tablewidth{0pt}
\tablecaption{Narrow Emission Lines on Day +2}
\tablehead{
  \colhead{} &
  \colhead{EW} &
  \colhead{Flux$^{a}$}\\
  \colhead{Line} &
  \colhead{($-$\AA)} &
  \colhead{($10^{-15}$ erg cm$^{-2}$ s$^{-1}$)}
}
\startdata
\ion{He}{2} $\lambda$4200         &  $2.6 \pm 0.8$  &  $6.8 \pm 2.0$ \\
\ion{He}{2} $\lambda$4686         &  $1.2 \pm 0.1$  &  $3.0 \pm 0.3$ \\
H$\beta$ $\lambda$4861            &  $0.7 \pm 0.2$  &  $1.6 \pm 0.4$ \\
\ion{C}{4} $\lambda\lambda$5805   &  $2.3 \pm 0.2$  &  $3.8 \pm 0.3$ \\
H$\alpha$ $\lambda$6563           &  $3.1 \pm 0.1$  &  $4.3 \pm 0.1$ \\
\enddata
\tablecomments{Equivelent widths and line fluxes measured via gaussian
fits to the continuum subtracted HET spectra on day +2 (combined east
and west tracks). Error estimates do not include systematics.}
\tablenotetext{a}{The flux scale was calibrated using the {\it Swift}
$V$-band photometry.}
\label{narrow_lines}
\end{deluxetable}


\begin{deluxetable}{rrrrrrrrrrrrrr}
\tablewidth{0pt}
\tablecaption{Measured Line Minima Velocities}
\tabletypesize{\scriptsize}
\tablehead{
  \colhead{} &
  \multicolumn{13}{c}{Phase (days)} \\
  \colhead{Line} &
  \colhead{2.1} &
  \colhead{3.4} &
  \colhead{6.3} &
  \colhead{8.1} &
  \colhead{10.1} &
  \colhead{12.1} &
  \colhead{16.1} &
  \colhead{21.3} &
  \colhead{25.3} &
  \colhead{33.3} &
  \colhead{42.2} &
  \colhead{57.2} &
  \colhead{73.1} 
}
\startdata
       H$\gamma$ $\lambda$4340 &  \nodata &    13.91 &    12.19 &    12.19 &    12.46 &    11.28 &     9.36 &     9.89 &     8.57 &     7.41 &     7.39 &     6.29 &     6.27 \\
     \ion{Fe}{2} $\lambda$4629 &  \nodata &  \nodata &  \nodata &  \nodata &  \nodata &  \nodata &  \nodata &  \nodata &  \nodata &  \nodata &     4.58 &     3.91 &     3.27 \\
     \ion{Sc}{2} $\lambda$4670 &  \nodata &  \nodata &  \nodata &  \nodata &  \nodata &  \nodata &  \nodata &  \nodata &  \nodata &  \nodata &  \nodata &     3.89 &     3.43 \\
     \ion{He}{2} $\lambda$4686 &    18.33 &    16.71 &    13.53 &    12.59 &  \nodata &  \nodata &  \nodata &  \nodata &  \nodata &  \nodata &  \nodata &  \nodata &  \nodata \\
        H$\beta$ $\lambda$4861 &  \nodata &    13.12 &    11.16 &    11.66 &    10.76 &    10.61 &    10.21 &     9.42 &     8.17 &     6.50 &     6.04 &     5.61 &     5.28 \\
     \ion{Fe}{2} $\lambda$4924 &  \nodata &  \nodata &  \nodata &  \nodata &  \nodata &  \nodata &  \nodata &     7.95 &     6.61 &     5.59 &     4.78 &     4.04 &     3.40 \\
     \ion{Fe}{2} $\lambda$5018 &  \nodata &  \nodata &  \nodata &  \nodata &  \nodata &  \nodata &    10.23 &     7.80 &     6.83 &     5.94 &     4.89 &     4.19 &     3.62 \\
     \ion{Fe}{2} $\lambda$5169 &  \nodata &  \nodata &  \nodata &  \nodata &    12.20 &    11.13 &     9.77 &     8.04 &     6.86 &     5.60 &     5.00 &     4.40 &     3.95 \\
     \ion{Fe}{2} $\lambda$5276 &  \nodata &  \nodata &  \nodata &  \nodata &  \nodata &  \nodata &  \nodata &  \nodata &  \nodata &  \nodata &  \nodata &     3.70 &     3.36 \\
     \ion{Fe}{2} $\lambda$5318 &  \nodata &  \nodata &  \nodata &  \nodata &  \nodata &  \nodata &  \nodata &     8.13 &     6.74 &     5.57 &     4.62 &     3.94 &     3.20 \\
     \ion{Sc}{2} $\lambda$5533 &  \nodata &  \nodata &  \nodata &  \nodata &  \nodata &  \nodata &  \nodata &     7.75 &     6.84 &     6.02 &     4.96 &     4.00 &     3.43 \\
     \ion{Fe}{2} $\lambda$5666 &  \nodata &  \nodata &  \nodata &  \nodata &  \nodata &  \nodata &  \nodata &     7.22 &     6.80 &     5.68 &     4.94 &     3.94 &     3.55 \\
     \ion{He}{1} $\lambda$5876 &    15.77 &    13.93 &    11.41 &    11.63 &    10.53 &    12.15 &    11.34 &    10.91 &    10.46 &    10.20 &    10.07 &     9.91 &     9.81 \\
     \ion{Na}{1} $\lambda$5892 &  \nodata &  \nodata &  \nodata &  \nodata &  \nodata &     9.36 &     8.15 &     6.67 &     6.57 &     5.90 &     5.25 &     4.98 &     5.11 \\
     \ion{Si}{2} $\lambda$6355 &  \nodata &  \nodata &  \nodata &  \nodata &    11.89 &     8.85 &     7.67 &     6.76 &     6.04 &  \nodata &  \nodata &  \nodata &  \nodata \\
       H$\alpha$ $\lambda$6563 &    15.38 &    14.48 &    12.77 &    11.40 &    11.01 &    10.65 &    10.38 &     9.87 &     9.23 &     8.45 &     7.74 &     6.95 &     6.82 \\
      \ion{O}{1} $\lambda$7773 &  \nodata &  \nodata &  \nodata &    13.00 &    12.71 &     9.69 &     9.04 &     7.32 &     6.35 &     5.51 &     4.64 &     3.90 &     3.40 \\
     \ion{Ca}{2} $\lambda$8520 &  \nodata &  \nodata &  \nodata &  \nodata &  \nodata &  \nodata &     7.91 &     8.30 &     7.93 &     7.09 &     6.65 &     6.05 &     5.61 \\
     \ion{Ca}{2} $\lambda$8662 &  \nodata &  \nodata &  \nodata &  \nodata &  \nodata &  \nodata &  \nodata &     8.87 &     7.69 &     6.45 &     5.87 &     5.10 &     4.61 \\
      P$\delta$ $\lambda$10049 &  \nodata &  \nodata &  \nodata &  \nodata &  \nodata &  \nodata &  \nodata &  \nodata &  \nodata &  \nodata &     5.69 &     5.01 &     4.06 \\
\enddata
\tablecomments{Velocities measured from the FT smoothed minima of selected
absorption features in units of $10^{3}$ km s$^{-1}$. Phases are relative to
2006 April 9.0}
\label{vel_table}
\end{deluxetable}


\end{document}